\RequirePackage{fix-cm}
\documentclass[smallextended, onecolumn]{svjour3}       
\smartqed  
\usepackage{graphicx}
\usepackage{amsmath}
\usepackage{booktabs} 
\usepackage{latexsym}
\usepackage[numbers]{natbib}
\usepackage{hyperref}
\usepackage{acro}
\usepackage{multirow}
\def\tsc#1{\csdef{#1}{\textsc{\lowercase{#1}}\xspace}}
\tsc{WGM}
\tsc{QE}
\journalname{Lifetime Data Analysis}

\DeclareAcronym{RUL}{
  short=RUL,
  long=Remaining Useful Life,
}
\DeclareAcronym{PHM}{
  short=PHM,
  long=Prognostic Health Management,
}
\DeclareAcronym{PoF}{
  short=PoF,
  long=Physics of Failure,
}
\DeclareAcronym{MFPCA}{
  short=MFPCA,
  long=Multivariate Functional Principal Component Analysis,
}
\DeclareAcronym{FDA}{
  short=FDA,
  long=Functional Data Analysis,
}
\DeclareAcronym{NASA}{
  short=NASA,
  long=National Aeronautics and Space Administration,
}
\DeclareAcronym{MFPCs}{
  short=MFPCs,
  long=Multivariate Functional Principal Components,
}

\DeclareAcronym{ML}{
  short=ML,
  long=Machine Learning,
}

\begin{document}

\title{Health Prognostics in Multi-sensor Systems Based on Multivariate Functional Data Analysis
}


\author{Cevahir Yildirim  \and Alba M. Franco \and Rosa E. Lillo \thanks{The research of the authors was partially supported by research grants and projects PDC2022-133359-I00, PID2022-137243OB-I00 and PID2022-137050NB-I00 of the Spanish Ministry of Science and Innovation.}
}


\institute{Cevahir Yildirim \at
              uc3m - Santander Big Data Institute (IBiDat), Spain \\
              \email{cevahir.yildirim@alumnos.uc3m.es}             \\
              ORCID: 0000-0002-8880-0955
           \and
           Alba M. Franco Pereira \at
              Interdisciplinary Mathematics Institute (IMI), UCM, Spain \\
              \email{albfranc@ucm.es} \\
              ORCID: 0000-0002-7480-1770
           \and
           Rosa E. Lillo \at
              uc3m - Santander Big Data Institute (IBiDat), Spain \\
              \email{rosaelvira.lillo@uc3m.es}  \\
              ORCID: 0000-0003-0802-4691
}

\date{Received: Feb 2025 / Accepted: date}

\maketitle

\begin{abstract}
Recent developments in big data analysis, machine learning, Industry 4.0, and IoT applications have enabled the monitoring and processing of multi-sensor data collected from systems, allowing for the prediction of the ``Remaining Useful Life'' (RUL) of system components. Particularly in the aviation industry, Prognostic Health Management (PHM) has become one of the most important practices for ensuring reliability and safety. Not only is the accuracy of RUL prediction important, but the implementability of techniques, domain adaptability, and interpretability of system degradation behaviors have also become essential. In this paper, the data collected from the multi-sensor environment of complex systems are processed using a Functional Data Analysis (FDA) approach to predict when the systems will fail and to understand and interpret the systems' life cycles. The approach is applied to the C-MAPSS datasets shared by \ac{NASA}, and the behaviors of the sensors in aircraft engine failures are adaptively modeled with Multivariate Functional Principal Component Analysis (MFPCA). While the results indicate that the proposed method predicts the RUL competitively compared to other methods in the literature, it also demonstrates how multivariate Functional Data Analysis is useful for interpretability in prognostic studies within multi-sensor environments.

\end{abstract}

\newpage
\section{Introduction}\label{Introduction}
\ac{PHM} aims to increase the reliability of systems and reduce downtime \cite{li2021hierarchical}. Accurate RUL prediction, which shows how long the system can operate without failure, plays a critical role in electromechanical systems (e.g., aircraft engines). In PHM, converting planned maintenance to on-condition maintenance is valuable for reducing maintenance costs and avoiding catastrophic failures \cite{xia2021lstm}. Each piece of equipment is an asset to any organization, and predicting failures is crucial to preventing shutdowns or performance degradation. Comprehensive monitoring of devices and systems helps predict when maintenance is required to prevent failure \cite{anandan2022industrial}.

The existing approaches in health prognostic studies can be classified as model-based and data-driven approaches \cite{zhang2020aircraft}. In model-based approaches, a mathematical model describes the system's health degradation, which typically must be developed using ``Physics of Failure'' (PoF) characteristics before prediction. Several model-based approaches have been developed for RUL prediction, such as the particle filtering-based approach \cite{skordilis2019double} and the Markov model-based approach \cite{ALDAHIDI2016109}. While model-based approaches can enhance accuracy in RUL prediction, the simplifications and assumptions inherent in the adopted models may increase limitations in practical implementations. Recent advancements in data mining and health monitoring have increased interest in data-driven approaches. Data-driven approaches are particularly applicable in PHM for complex systems, such as aircraft engines, which may lack PoF knowledge but benefit from extensive real-life monitoring in a multi-sensor environment.

The common data-driven \ac{ML} approaches used for prognostics include artificial neural network-based models \cite{de2022alarm} and long short-term memory models \cite{huang2019bidirectional, al2019multimodal, xu2022global, xia2021lstm}, the Cox proportional hazard model \cite{zhu2021real}, support vector regression \cite{khelif2016direct}, other regression-based models \cite{lu2019aircraft}, and fuzzy logic \cite{li2019degradation}, among others. Machine learning methods are capable of addressing prognostic challenges in complex systems. However, certain uncertainties must be managed in engineering applications, such as the effects of manufacturing variations, applied maintenance, and differences in component quality among systems under prognostics. Additional studies have been conducted to estimate these variation points or change points in hazard functions \cite{gijbels2003estimation}. Managing uncertainties in long-term device prognostics remains challenging due to a lack of transparency. In contrast, to better address uncertainties in the degradation process, fitting probabilistic models using statistical approaches may be more effective for future predictions. As a result, statistical prognostics has recently gained prominence \cite{hanachi2018performance}. Regarding statistical approaches in health prognostics, regression-based models such as the multivariate regression splines method \cite{lasheras2015hybrid}, Bayesian hierarchical models \cite{zaidan2016gas}, and parametric $p$th-order polynomial models \cite{song2018statistical, liu2014integration} have been applied to predict future health conditions. Stochastic process models are also widely studied due to their ability to describe the degradation dynamics of systems. The Wiener process has been explored for engineering system prognostics \cite{zhang2018degradation}. The non-homogeneous Gamma process has been investigated under the assumption of non-decreasing (monotone) degradation \cite{le2016remaining}, while the inverse Gaussian process has been used to model time-varying degradation rates \cite{peng2017bayesian}. Additionally, an FPCA approach was studied in \cite{zhang2020aircraft}, where the RUL was effectively predicted using an FDA technique.

As mentioned earlier, multi-sensor data collected from systems are commonly used in the prognostics of complex systems. However, the relevance of each sensor to failure varies. The information obtained from each sensor may have different levels of influence on the system's failure behavior. Some sensor values may remain constant over time and be unrelated to degradation, while others may vary but be inconsistent and still unrelated to the degradation process. Additionally, some sensors may have higher noise levels or lower sensitivity than others. Analyzing too many irrelevant sensors is both time-consuming and costly, making the elimination of non-informative sensors essential. Proposed sensor elimination methods include logistic regression \cite{yu2017aircraft}, hypothesis testing \cite{chehade2019sensor}, and permutation entropy-based methods \cite{liu2017quantitative}. However, the existing literature primarily focuses on selecting sensors that exhibit a consistently decreasing or increasing trend over time across all engines while ignoring those that remain constant or behave inconsistently over time. In general, most statistical techniques fall within the scope of univariate degradation modeling. For instance, a univariate metric called the "Health Index" is constructed to create a univariate representation of informative multi-sensor data after eliminating non-informative sensors. Health index construction methods based on linear transformations have been proposed in \cite{wang2012generic, kim2019generic} and have been widely applied in various studies.

The fusion of remaining sensor data after eliminating non-informative sensors and the generation of a univariate metric are commonly used in engine prognostics across various statistical methods. When the degradation process can be described using a parametric model, such as a linear model \cite{kim2019generic}, quadratic model \cite{coble2011applying}, exponential model \cite{liu2014integration, fang2017multistream}, or Weibull models, analysis can be effectively conducted using parametric techniques. However, in many cases, the degradation pattern does not fit well within a parametric model \cite{volponi2014gas,hanachi2018performance}, necessitating the use of non-parametric techniques and machine learning approaches. 

Various approaches have been compared, and the challenges of machine learning-based RUL prognosis have been systematically reviewed in the literature. While most research has primarily focused on minimizing error rates, it has been argued that additional factors—such as interpretability, robustness, and domain adaptability—should also be considered for practical implementation \cite{vollert2021challenges}. Since maintenance involves costly repairs or part replacements, users demand transparent and interpretable RUL predictions. As a result, the field of interpretable machine learning is expected to gain increasing importance in RUL prediction and may become a decisive factor in its adoption. 

Furthermore, many developed models remain highly specific to particular applications. More flexible approaches that emphasize domain adaptability and broader applicability across different systems have become crucial. On one hand, directly utilizing multivariate sensor data instead of generating a univariate index can help prevent information loss caused by artificial metric generation. Consequently, the interpretability of raw sensor data is often more beneficial than univariate health indexes or other extracted features, which may omit critical information. On the other hand, since feature extraction methods and univariate health indexes vary across applications, a more adaptable approach suitable for diverse scenarios is required for practical use. To address these challenges, this paper proposes a FDA method using MFPCA to predict and interpret raw multi-sensor failure behavior and estimate RUL. 

The contributions of this work are as follows. A novel MFPCA-based FDA approach is proposed to address the challenges of PHM studies. This approach enables the direct use of raw multi-sensor data, eliminating the need for feature extraction or univariate health indexes. 

A comparative analysis of the proposed MFPCA approach and the univariate FPCA method, based on Root Mean Square Error (RMSE) in RUL prediction, demonstrates superior performance. Additionally, this work tackles a critical challenge in PHM studies: interpreting the predicted RUL path. A dynamic, time-varying RUL interpretation system is introduced, enhancing both interpretability and implementability—two major concerns in PHM research.

To validate the proposed technique, we use the C-MAPSS aircraft engine degradation dataset, a well-known benchmark published by NASA as part of the 2008 PHM challenge \cite{saxena2008turbofan}. Over the years, various machine learning approaches have been applied to this dataset, making it a strong reference for comparison.

The paper is structured as follows: Section \ref{Background} details the proposed model structure, including the registration process, basis representation, functional data smoothing, MFPCA, Karhunen–Loève expansion, and similarity-based RUL prediction. Section \ref{Case Study} presents a case study using the C-MAPSS dataset, demonstrating the application of all methodological steps. Finally, Section \ref{Conclusions} provides conclusions and future research directions.

\section{Structure and the Background of the Proposed Method}\label{Background}

\begin{figure*}[t]
    \centering
    \includegraphics[width=15cm]{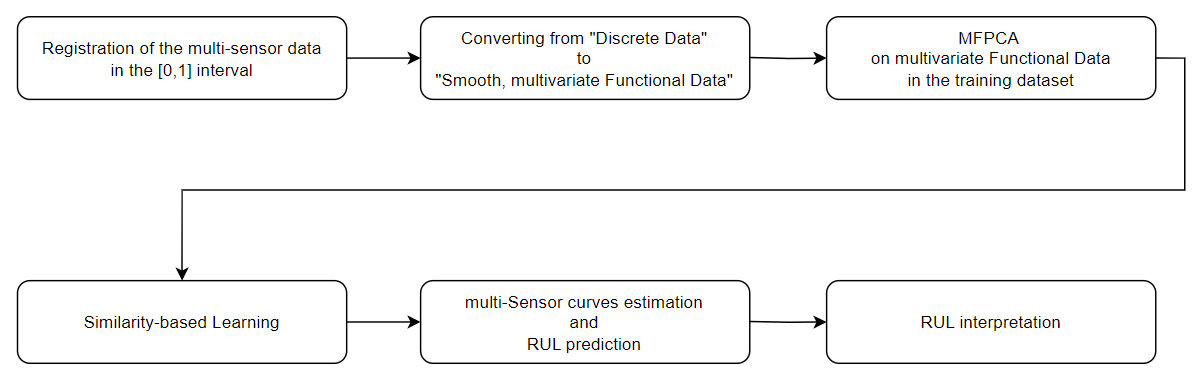}
    \caption{Proposed Health Prognostic Approach Based on Multivariate Functional Data Analysis in Multi-sensor Systems}
    \label{fig:steps1}
\end{figure*}

The structure (see Fig.\ref{fig:steps1}), preliminary information, and the basic tools of the proposed FDA approach are given in this section. The method provides an MFPCA-based learning approach, and the data becomes applicable to perform degradation behavior interpretation and similarity-based RUL prediction for each test system.
The multi-sensor data collected until failure in engineering systems can be considered as times-series multivariate functional data. The matrix in Table \ref{table1} represents the functions and vectors for each individual system and sensor observation set for these kinds of problems. Mathematically, the available data can be represented in a matrix where each cell ($i,j$) collects the evolution of sensor $j$ in system $i$, $i=1,\mathellipsis, n$, and $j=1,\mathellipsis,J$. We denote those functions as $X_{ij}(t)$. Now, in practice, the function $X_{ij}(t)$ is observed at discrete instants of time ($t_{i,1},\mathellipsis, t_{i,T_i}$) where $T_i$ is the failure time for $i$th system and this failure time can be different for each system. It shows that the available data can be modeled as a sample of size $n$ of multivariate functional data of dimension $J$. In this context, obtaining MFPCs presents several issues to handle. On the one hand, the cycle times are not defined in the same domain because the time to failure is different among cycles for each individual. On the other hand, we only have discrete observations of each $i$th system at a finite set of sensor values until the $T_{i-failure}$ point. To solve these problems, obtain interpretable data, and predict the RUL, we propose a novel FDA approach based on seven main steps given in Fig. \ref{fig:steps1}.

Since the proposed method is based on MFPCA, the fundemental tools of FDA must be introduced to enable direct analysis of the dataset observations. Background information on curve registration \cite{ramsay2005functional} is given in Section \ref{Registration}. Preliminary information of basis representation and smoothing is provided in Section \ref{prelim-smoothing}, MFPCA \cite{berrendero2011principal}, and Karhunen-Loève Expansion (KLE) \cite{karhunen1946spektraltheorie,loeve1945calcul} are briefly explained in Section \ref{MFPCAandKLE}. A similarity-based RUL prediction method, which is based on finding the sensor curves in the training group that are most similar to the sensors in the test group and making predictions accordingly, is explained in Section \ref{Similaritybased}.

\begin{table}[ht] 
\small
\caption{Functions and vectors for each individual system/sensor observation set} \label{table1}
\centering
    \begin{tabular}{ c| c c c}
  &\textbf{sensor 1} & \textbf{ $...$} &\textbf{sensor J}  \\  \hline
   &\multirow{3}{3em} { } &&    \\
\textbf{indiv. sys. 1}&  { \centering $X_{11}(t)$}  & $...$ &  { \centering $X_{1j}(t)$}  \\
& $X_{11}(t_{1,1})$ , $...$ , $X_{11}(t_{1,T_{1}})$  & & $X_{1j}(t_{1,1})$ , $...$ , $X_{1j}(t_{1,T_{1}})$  \\
& & &\\
   &\multirow{3}{3em}  &  &    \\
\textbf{\vdots}&   $\vdots$ & $\ddots$ &   $\vdots$\\
& & &\\
   &\multirow{3}{3em} { } & &    \\
\textbf{indiv. sys. n}& { \centering $X_{n1}(t)$} & $...$ &  { \centering $X_{nj}(t)$}  \\
&$X_{n1}(t_{n,1})$ , $...$ , $X_{n1}(t_{n,T_{n}})$  &  & $X_{nj}(t_{n,1})$ , $...$ , $X_{nj}(t_{n,T_{n}})$  \\
\end{tabular}
    \label{pdm}
\end{table} 

\subsection{Registration}\label{Registration}
In FDA, it is common for all curves to be defined within the same domain, which is not the case in Table \ref{table1}. Therefore, curve registration is required to normalize them to the interval [0,1] (see \cite{ramsay2005functional}, Chap. 7, for a detailed description). To achieve this, we propose transforming the domain $[0,T_{i-endpoint} ]$ of each $i\textsuperscript{th}$ using the function $t / T_{i-endpoint}$. Then, FDA methodologies will be performed on the synchronized curves given by
\begin{equation}
X_{i}^*(t)=X_{i}(t \cdot T_{i-endpoint}), \quad \text{for all} \quad  t  \in  [0,1].
\label{eq1}
\end{equation}
Therefore,
\begin{equation}
X_{i}(t)=X_{i}^*\Big( \frac{t}{T_{i-endpoint}} \Big), \quad \text{for all} \quad  t  \in  [0,T_{i-endpoint}]
\end{equation}
while $X_{i}$ is representing the original scale for the $i\textsuperscript{th}$ system, $X_{i}^*$ is its corresponding registered function where $i=1,\mathellipsis n$.
Doing this transformation, we warranty to have the same domain for the curves in the dataset. This way, for each curve, we have a new registered observation point in the interval $[0,1]$ given by
\begin{equation}
v_{ik}=\frac{t_{ik}}{T_{i-endpoint}}, \quad k=1,\mathellipsis, T_{i-endpoint}.
\end{equation}
where $v_{ik}$ is the new argument for the $i\textsuperscript{th}$ system at $k\textsuperscript{th}$ observation point.

This means that the sampling points at which the registered curves are observed differ not only in number but also in position.  As a result, the new curves have become unequally spaced. However, it is important to note that the registration process does not alter the shape of the curves, preserving valuable information for understanding and interpreting their behavior.

\subsection{Basis Representation and Smoothing} \label{prelim-smoothing}

As shown in Fig.\ref{fig:steps1}, the next step in FDA is reconstructing the sample in a functional form from its discrete observations. The most common approach is to express each sample curve as an expansion in terms of a basis of functions and estimate the basis coefficients using smoothing \cite{aguilera2013comparative}. Let $X_i(t), i=1,\mathellipsis,n$ be a sample function generated by a process $X(t)$. In practice, sample functions are observed at a finite set of time points ($t_{i,0},t_{i,1},\mathellipsis,t_{i, T_i}$ $\in$  $T$) for all $i = 1,\mathellipsis,n$. The sample information is then represented by the vectors, $x_i=(x_{i,0}, \mathellipsis, x_{i,T_i})$, where $x_{i,0}$ is the sensor observation of the $i$th system at time point 0, and $x_{i,T_i}$ is the $i$th system's sensor observation value at failure time for all $i = 1,\mathellipsis,n$. In this section, the sample paths are assumed to belong to a finite-dimension space generated by a basis ${\phi_{1}(t),\mathellipsis,\phi_B(t)}$, so that they are expressed as
\begin{equation} \label{eqXij}
X_i(t) = \displaystyle\sum_{b=1}^{B} c_{ib}\phi_b(t), \quad i=1,\mathellipsis, n.
\end{equation}
where $c_{ib}$ are the coefficients of the basis functions for the i\textsuperscript{th} individual observation set and the $b^{th}$ basis function with $B$ number of basis functions. The function of $X_i(t)$ which is a linear combination of ${\phi_{1}(t),\mathellipsis,\phi_B(t)} : t \in T$, is called a functional data \cite{ramsay2005functional}. The selection of the basis and its dimension $B$ is crucial and must be done according to the characteristics of the curves. Useful basis systems are Fourier basis for periodic data, Bspline basis for non-periodic smooth data with continuous derivatives up to certain order, and wavelet basis for data with a strong local behavior whose derivatives are not required \cite{aguilera2010using,ramsayapplied}.

In this paper, cubic B-splines with a discrete roughness penalty are used to reconstruct the sensor curves, with the dimension of the B-spline basis selected through cross-validation. Since the sensor curves are sufficiently smooth to allow for an accurate spline approximation, an alternative approach could be to use a P-spline basis \cite{aguilera2019stochastic}.

A B-spline basis of order 4 (cubic splines) defines a space of splines of the same degree, with each segment represented by a distinct cubic polynomial function. Cubic B-splines are the most commonly used in practice because they ensure continuity of both the first and second derivatives. According to Šulejic \cite{sulejic2011} , a $n$th-degree Bspline curve (order of $n + 1$) is defined by
\begin{equation}
 C(t)= \displaystyle\sum_{k=1}^{\infty} B_{k,n}(t)P_k,  
\end{equation}
where $P_k$ are the control points. The Bspline curve $C(t)$ is constructed from $n$th-degree basis functions $B_{k,n}(t)$ defined by recurrence on the inferior degree. Theoretically, the number of control points and spline functions can be infinite, but it should be limited at some point in practice.

We define the knot vector as $T=(t_0,\mathellipsis,t_m)$ where $T$ is a non-decreasing sequence of real numbers satisfying $t_k\leq t_{k+1}$ for $k =0,\mathellipsis,m-1$. The values $t_k$ are referred to as knots. The $k$th Bspline basis function of  degree $n$ is defined by the Cox-de Boor recursion formula \cite{de1978practical}:
\begin{equation}
B_{k,0}(t) = \left\{ \begin{array}{ll}
1, & \textrm{ if \quad $t_k\leq t \leq t_{k+1}$},\\
0, & \textrm{otherwise},
\end{array} \right.
\end{equation}
with
\begin{equation} \label{cubicsplines}
B_{k,n}(t) = \frac{t-t_k}{ t_{k+n}-t_k } B_{k,n-1}(t) + \frac{t_{k+n+1}-t}{t_{k+n+1}-t_{k+1}}B_{k+1,n-1}(t)
\end{equation}

In general, the knots of a Bspline are equally spaced, and their number is chosen to balance data fit and computational efficiency. Several key factors influence B-spline fitting, including  the smoothing parameter, the order of the penalty, the degree of the B-spline basis, and the number of knots. A common and effective approach for selecting these parameters is to use cross-validation to determine the smoothing parameter, apply a quadratic penalty, use cubic splines, and set one knot for every four to five observations, with a maximum of approximately 20 knots (see \cite{aguilera2013comparative} for a comparative study on the performance of regression splines, smoothing splines, and P-splines in both simulated and real-life datasets).

To ensure consistency across all $n$ sample paths, we propose selecting a single smoothing parameter by minimizing the mean leave-one-out cross-validation error across the entire set of sample curves. The leave-one-out cross-validation (CV) method consists of selecting the smoothing parameter $\lambda$ that minimizes
\begin{equation}
CV(\lambda)=\frac{1}{n}\displaystyle\sum_{i=1}^n CV_i(\lambda),
\end{equation}
where
\begin{equation}
CV_i(\lambda)=\sqrt{\displaystyle\sum_{k=1}^{Ki} (X_{ik}-\widehat{X}_{ik}^{-k})^2 / (K_i + 1)},
\end{equation}
with $\widehat{X}_{ik}^{-k}$ being the values of the $i$th sample path estimated at time $t_{ik}$ avoiding the $k$th time point in the iterative estimation process. A computationally simplest approach very used in the literature about smoothing splines is generalized cross-validation (GCV) \cite{craven1979estimating}. The GCV method consists of selecting $\lambda$, which minimizes
\begin{equation}
GCV(\lambda)=\frac{1}{n}\displaystyle\sum_{i=1}^n GCV_i(\lambda),
\label{eqgcv}
\end{equation}
where
\begin{equation}
GCV_i(\lambda)=\frac{(K_i+1) MSE_i(\lambda)}{[trace(X-H_i(\lambda))]^2},
\end{equation}
with $MSE_i(\lambda)= \frac{1}{n}\sum_{k=1}^{k_i}(X_{ik}-\widehat{X}_{ik})^2$ and $H_i(\lambda)=\phi({\phi_i^{'}} \phi_i+\lambda P_d)^{-1}{{\phi_i}^{'}}, \phi_i$ being the B-spline basis evaluated at the observation knots and $P_d$ the discrete penalty matrix. 

\subsection{MFPCA and Karhunen–Loève Expansion} \label{MFPCAandKLE}
The main goal of principal component analysis in a multivariate setting is to reduce the dimensionality of a dataset containing a large number of correlated variables while preserving as much of the variation as possible \cite{gorecki2018selected}. This is achieved by transforming the original variables into a new set of variables known as principal components, which are ordered such that the first few retain most of the variation present in the original dataset. It is well known in the statistics that if we observe a $j$-dimensional random vector $\textbf{X}=(X_1, X_2,\mathellipsis, X_j)^{'}  \in {R}^j $ The first step is to find a linear combination $U_1=u_{11}X_1 +u_{12}X_2 +\mathellipsis+u_{1j}X_j= u^{'}_1X$ that maximizes the variance. This new variable $U_1$ is called the first principal component. Next, a second linear combination, $U_2 = u^{'}_2X$, is determined, ensuring that it is uncorrelated with $U_1$ while still capturing the maximum remaining variance. This process continues iteratively. At the $k$th stage, the principal component $U_k = u^{'}_kX$ is identified as the linear combination with the highest variance among those uncorrelated with the first $k-1$ principal components \cite{jolliffe2002principal}.

Principal component analysis for multivariate functional data was first proposed by Ramsay and Silverman (2005) \cite{ramsay2005functional}  and Berrendero et al. (2011) \cite{berrendero2011principal}. The functional verison of PCA is referred to as FPCA. FPCA offers an insightful way to examine the variability structure within the variance-covariance operator for one-dimensional functional data \cite{gorecki2012functional}. Let us assume we have a sample of curves representing $p$ different variables across in $n$ individuals. Therefore, we can represent the data with FPCA in a multivariate sense. Now, in the multivariate case, we have $\textbf{X(t)}= {(X_{1}(t), X_{2}(t),\mathellipsis,X_{p}(t))}^{'}$ with $t \in I $, and assume that $\textbf{X} \in L_{j}^2(I) $ is a Hilbert space of square-integrable functions on the interval $I$, and $j=1,\mathellipsis,p$. For $t,u \in I$, we define the covariance matrix $C(t,u)$ in the $j$th domain as 
\begin{equation}
C_j(t,u)=Cov\big(X_j(t),X_j(u)\big), \quad j=1,\mathellipsis,p \quad t,u \in I.
\end{equation}
As noted in \cite{ramsay2005functional}, a suitable inner product is the basis of all approaches for principal component analysis. For functions $X=(X_{1},\mathellipsis,X_{p})^{'}$ with $X_{p} \in L_{j}^2(I)$ and $t \in I $, the inner product can be expressed as
\begin{equation}
<X,Y>= \sum_{j=1}^p \int_{I_{j}} X_{j}(t)Y_{j}(t)dt, \quad j=1,\mathellipsis,p \quad t \in I, \end{equation}
where the covariance operator of $X(t)$ is a positive auto-adjoint compact operator defined by
\begin{equation}
C\big(X\big)_{j}(t)=\sum_{j=1}^p \int_{I_{j}} C_j(t,u)X_{j}(u)du, \quad j=1,\mathellipsis,p \quad t \in I.
\end{equation}
Then, the spectral representation of $C$ provides the following orthogonal decomposition of the process, known as Karhunen–Loeve orthogonal expansion \cite{karhunen1946spektraltheorie,loeve1945calcul}
\begin{equation}
X_{j}(t)=\mu_{j}(t) + \displaystyle\sum_{k=1}^\infty f_{jk}(t) \xi_{jk}, \quad j=0,\mathellipsis,p \quad t \in I
\end{equation}
where $f_{jk}(t)$ are the orthonormal family of eigenfunctions of the covariance operator $C$ associated with its decreasing sequence of non null eigenvalues $\lambda_k$ that is
\begin{equation} \label{eqeigenvalues}
C_j\big(f_{j}(t)\big)=\int_{I_j} C_j(t,u)f_{j}(u)du = \lambda_{jk} f_{jk}(t), \quad t \in I.
\end{equation}

Similarly, $f_{jk}$ is called the $k$th principal weight function or harmonic factor for the $j$th domain. Taking into account that the total variance of $X_{j}(t)$ given by
\begin{equation} \label{eqtotalvariance}
    V_{j}=\int_{I_j} C_j(t,t)dt = \displaystyle\sum_{k=1}^\infty \lambda_{jk}, \quad j=1,\mathellipsis,p,
\end{equation}
then the ratio $\lambda_{jk}/V_{j}$ represents the variance explained by the $k$th functional principal component. In addition, the series truncated in the $q$th term is the best approximation of the process (in the least-squares sense) by a sum of $q$ quasi-deterministic terms \cite{saporta1981methodes}. Therefore the process admits the following multivariate functional principal component reconstruction in terms of the first $q$ principal components so that the sum of the variances explained by them is as close as possible to one.
\begin{equation} \label{eqxqu}
X_{ij}^{q}(t)=\mu_{j}(t) + \displaystyle\sum_{k=1}^q f_{jk}(t)\xi_{ijk}, \quad t \in I,
\end{equation}
where $\xi_{ijk}$ is defined as the family of uncorrelated zero-mean random variables, which can be defined in the multivariate sense by
\begin{equation} \label{eqksi}
\xi_{ijk}=\int_I f_{ijk}\big(X_{ij}(t) - \mu_{j}(t)\big)dt, \quad t \in I.
\end{equation}
The random variable $\xi_{ijk}$ is called the $k$th principal component and has the maximum variance $\lambda_k$ out of all the generalized linear combinations of $X_{ij}(t)$ which is the function of the $i$th individual at $j$th domain.

\subsection{Similarity-based Prediction for Degradation Behavior of Sensors and RUL} \label{Similaritybased}

The proposed method assumes that similar systems will exhibit comparable degradation behaviors. These behaviors are learned through MFPCA. In MFPCA analysis, the first few principal components, which explain a significant percentage of the total variance, provide valuable information about the functional curve behavior of the data over time. FDA practitioners use score plots and score distributions to interpret the MFPC scores effectively. A key focus is to determine whether individuals can be classified based on the first few MFPCs. If classification is feasible using these components, the next step is to qualitatively or quantitatively interpret the differences between individuals.

In multi-sensor PHM datasets, the first few functional principal components are expected to separate the dataset into several groups based on their MFPC scores. When a difference is observed between MFPC score groups in the training dataset, the next step is to identify the same difference in the test dataset. Once the MFPC scores are obtained and the dataset is grouped, we calculate the distance between a test system's sensor curve and training system curves with similar MFPC score groups. We will use and recommend the Euclidean ($L^2$) distance as it is the natural measure between two points in Euclidean space and represents the length of the line segment connecting two points. In case of multivariate distancing, it is advisable to normalize the component time series, as nonnormalized time series with relatively high values could dominate the distance measure. Normalizing the $n$-dimensional data before multivariate $L^2$ distance calculation with
\begin{equation} \label{eq:distance1}
u^*_{ij}(t)=\frac{X^*_{ij}(t)}{\bar{X}^*_{j}(t)} \quad i=1,\mathellipsis,n    \quad  j=1,\mathellipsis,J
\end{equation} 
where $X^*_{ij}$ is the sensor value of the $i$th {system and the $j$th sensor at time $t$, $\bar{X}^*_{j}$ is the mean of the $j$th sensor, and $J$ is the number of sensors. The distance between the two engine curves in a multivariate sense is expressed as
\begin{equation} \label{eq:distance2}
d_{x,y}(t)=\sqrt{\displaystyle\sum_{j=1}^{J} (x_j(t)-y_j(t))^2}, \quad t\in[0,1]
\end{equation} 
where $\sqrt{(x_j(t)-y_j(t))^2}$ is the length between the curve $x$ and curve $y$ in one dimension, and $J$ is the number of sensors. Using the obtained multivariate distance data, not only can the RUL of each system be estimated, but the multi-sensor curve up to the predicted point of failure can also be estimated by calculating the mean or median of the sensor values from similar training system curves at the relevant time points. RUL and multi-sensor curve predictions will be performed for the remaining cycles of all test systems. The mean and median failure times of the most similar training systems will be considered separately. In addition to RUL prediction, multi-sensor curve estimation along RUL is is crucial for the interpretability of RUL. One of the key advantages of FDA is its ability to utilize derivatives of functions. In FDA studies, it is common to consider not only the original functions but also their first and second derivatives, which is expected to provide valuable insights in PHM studies as well. Interpreting sensor curves will enable us to understand data in terms of maintenance needs, maintenance alarm points, degradation behaviors of different MFPC groups, and degradation behaviors of similar systems.

\section{Case Study}\label{Case Study} 
\makeatletter
\renewcommand{\fnum@figure}{\textbf{Fig.} \thefigure}
\makeatother
\begin{figure*}[b]
		\includegraphics[width=5.5cm]{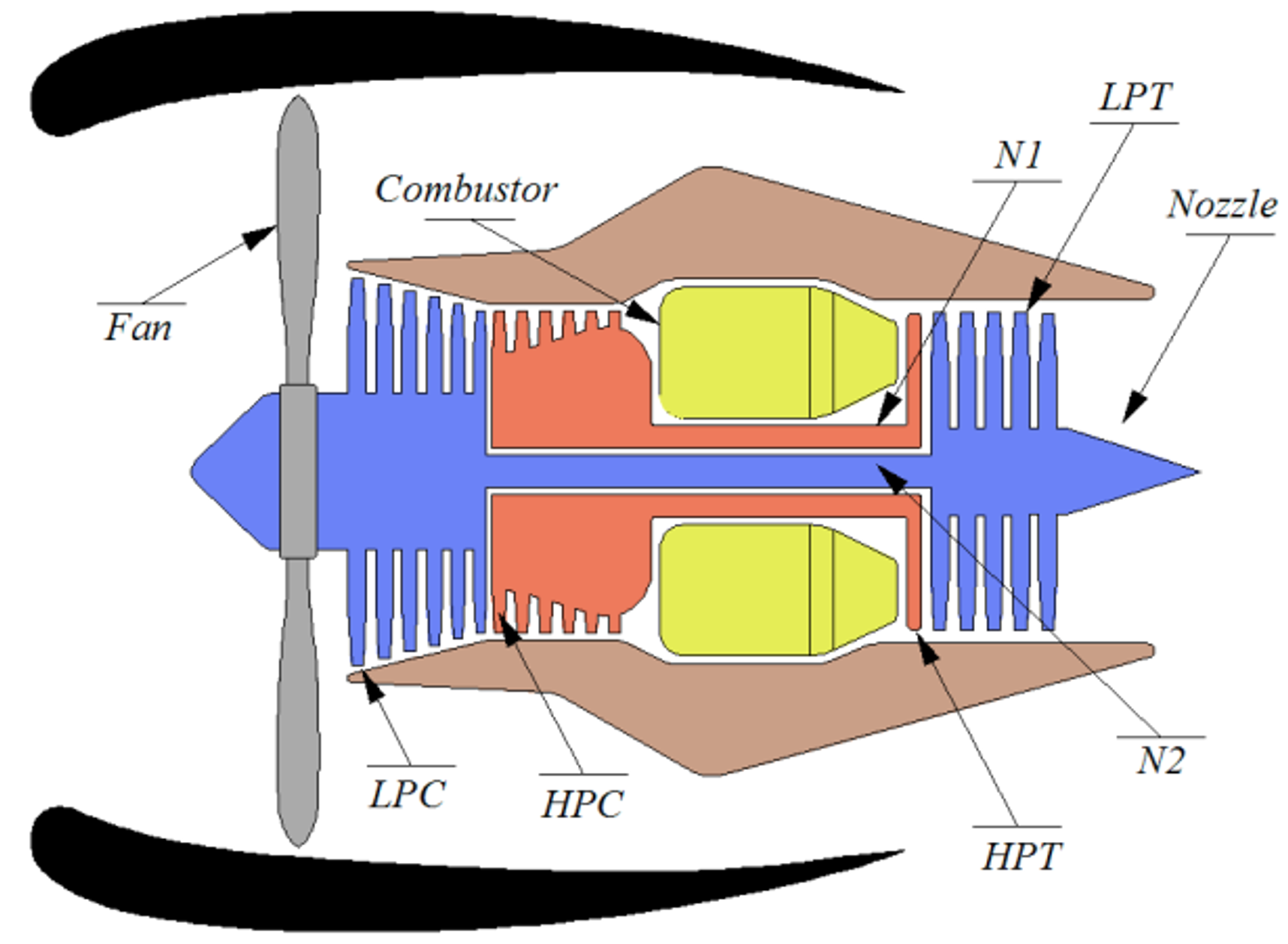}
	  \caption{Aircraft Gas Turbine Engine Sections}\label{fig1}
\end{figure*}
The sensor data of the aircraft gas turbine engine degradation is generated from the commercial modular aero-propulsion system simulation (C-MAPSS) developed at NASA \cite{saxena2008damage} and published online for research investigations. The data can be found and downloaded from NASA data repository \cite{saxena2008turbofan}. The chosen dataset FD001, contains 100 training and 100 test data for only one fault mode, which is the degradation in the ``HPC'' section shown in Fig.\ref{fig1}. datasets consist of multiple multivariate time series and each time series comes from a different engine. Each engine operates with varying degrees of initial wear and manufacturing variation unknown to the user. This wear and variation are considered normal. In other words, it is not considered a fault condition. Engines always run normally at the beginning of the series and fail at some point in the series. In the training set, sensor values are given until the failure, while in the test set, the time series ends at a point before the system failure. A vector of actual RUL values for the test data is also provided.

The sensor measurements for each training engine were collected until failure. Each time series signal represents a different degradation instance of the dynamic simulation of the same engine population and consists of multi-sensor measurements. For each cycle of a degradation instance, 21 sensor measurements were recorded as listed in Table \ref{tbl1}. After a rough screening of the 21 sensors, it is shown that the sensors T2, P2, P15, epr, farB, htBleed, Nfdmd, and PCNfR$_{dmd}$, and Nf$_{dmd}$, NRf are constant over time and excluded. Nc, and NRc are very inconsistent over time because the value of the sensor is sometimes increasing, sometimes decreasing, and sometimes constant during life cycles. These inconsistent sensors will be considered non-informative. In previous studies using the C-MAPSS data, including \cite{le2016remaining,wang2019data,zhang2020aircraft}, the above sensors were also eliminated as non-informative sensors. Basically, constant, binary, and inconsistent sensors were detected and eliminated. The remaining sensors selected as informative sensors provide nine-dimensional data for multivariate FDA (Fig.\ref{fig:trainsensors}.) As can be seen from the figure, some of the informative sensors tend to decrease over time, while others tend to increase. Although these engines were simulated in a single operating condition and single failure mode, the engine with multiple operating conditions and failure modes can also be analyzed using the proposed method with the necessary procedures as in \cite{le2016remaining,singh2019novel}. The aim of the study is to estimate the future of the sensor curves and the RUL of each engine in the "Test dataset", and to obtain interpretable multivariate functional data for each individual engine. Additionally, a practical study in the field of RUL Prediction and PHM is carried out and the use of a new, multivariate FDA approach is proposed.

\begin{table}[ht]
\caption{C-MAPSS Data Sensor Details}\label{tbl1}
\begin{tabular}{lll}
\hline\noalign{\smallskip}
Sensor & Description & Units\\  
\noalign{\smallskip}\hline\noalign{\smallskip}
T2 & Total temperature at fan inlet & °R\\
T24 & Total temperature at LPC outlet  & °R\\
T30 & Total temperature at HPC outlet  & °R\\
T50 & Total Total temperature at LPT outlet  & °R\\
P2 & Pressure at fan inlet  & psia\\
P15 & Total pressure in bypass-duct & psia\\
P30 & Total pressure at HPC outlet  & psia\\
Nf & Physical fan speed  & rpm\\
Nc & Physical core speed  &rpm\\
epr & Engine pressure ratio (P50/P2)  & -\\
Ps30 & Static pressure at HPC outlet  & psia\\
phi & Ratio of fuel flow to Ps30  & pps/psi\\
NRf & Corrected fan speed  & rpm\\
NRc & Corrected core speed  &rpm\\
BPR & Bypass Ratio  & -\\
farB & Burner fuel-air ratio  & -\\
htBleed & Bleed Enthalpy  & -\\
$Nf_{dmd}$ & Demanded fan speed  & rpm\\
$PCNfR_{dmd}$ & Total temperature at fan inlet & rpm\\
W31 & HPT coolant bleed   & lbm/s\\
W32 & LPT coolant bleed  & lbm/s\\
\noalign{\smallskip}\hline
\end{tabular}
\end{table}

The proposed multivariate FDA approach, shown in Fig \ref{fig:steps1}. is applied to the dataset. The matrix presented in Table \ref{table1} can represent the functions and vectors for each set of engine and sensor observations for the case study. Mathematically, the available data can be represented in a matrix where each cell ($i,j$) collects the evolution of sensor $j$ in engine $i$, $i=1,\mathellipsis, n$, and $j=1,\mathellipsis, J$. We denote those functions as $X_{ij}(t)$. In practice, the function $X_{ij}(t)$ is observed at discrete instants of time ($t_{i,1},\mathellipsis, t_{i, T_i}$) where $T_i$ is the failure time for engine $i$ ,which can vary across engines. Table \ref{table1} shows that the current data can be modeled as an $n$-dimensional sample of multivariate functional data of dimension $J$. However, before applying any known methodology in multivariate FDA, the data must undergo preprocessing. In FDA, it is common for all the functions involved in a study to have the same domain. Therefore, for the FDA approach to be applied, all observations for each engine must be registered at the same interval (see Section \ref{Registration}). Additionally, the engine/sensor vectors need to be converted into smooth functions to be analyzed within the FDA framework (see Section \ref{prelim-smoothing}). After this pre-processing, the data in the case study will be available for MFPCA analysis (see Section \ref{MFPCAandKLE}). To estimate the \ac{MFPCs}, we have a sample of functional data consisting of $J \times n$ sensor curves denoted by
\begin{equation} \label{eqdataexplained}
X_{ij} (t) \quad i=1,\mathellipsis,n;\quad j=1,\mathellipsis,J \quad t\in[0,T_{i-failure}]    
\end{equation}
where $T_{i-failure}$ is the failure time in terms of cycle time. $X_{ij}$ is the observation of $j$th sensor of the $i$th engine 
at a finite set of sensor values until the $T_{i-failure}$ point. Each curve $X_{ij}(t)$ is observed at $t_i$ equally spaced sampling points. Each curve of $X_{ij}(t)$ has a different domain as they have different failure times $T_{i-failure}$. Ultimately, after learning the sensor behaviors up to the failure point for each engine, it will be possible to predict similarity-based sensor curves and the RUL. (see Section \ref{Similaritybased}).

\begin{figure}[t]
    \centering
    \includegraphics[width=15cm]{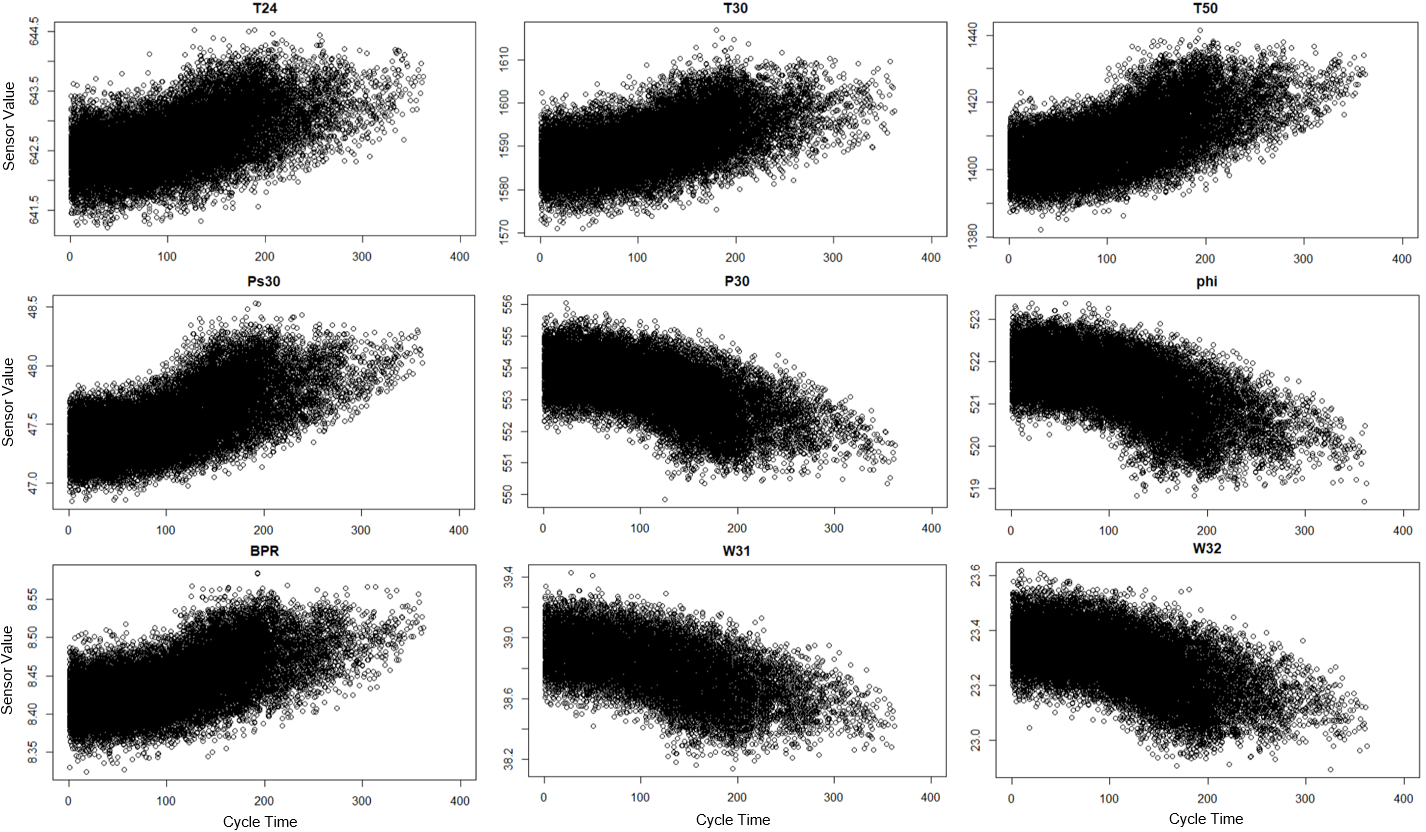}
    \caption{The observations of 9 informative sensors until failure for 100 training engines }
    \label{fig:trainsensors}
\end{figure}

\subsection{Registration of the Multivariate Sensor Data}\label{Registration of sensor curves}

The first pre-processing step is to register the sensor data within the [0,1] interval to ensure that all the observations defined in the same interval. Using the equation (\ref{eq1}), the representation of the dataset in the original scale (\ref{eqdataexplained}) will turn into
\begin{equation} 
X^*_{ij} (t) \quad i=1,\mathellipsis,n,\quad j=1,\mathellipsis,J, \quad t\in[0,1],    
\end{equation}
in the registered interval while $X_{ij}^*$ representing the registered function for $i$th engine and $j$th sensor where $i=1,\mathellipsis,n$ and $j=1,\mathellipsis,J$ explained in the matrix (Table. \ref{table1}) defined in Section \ref{Background}. Moreover, all the engines start at 0 and fail at 1 on the new scale. In our particular case, we have $n=100, J=9$, and $I$ in [0,1] because there are 100 independent training engines and 9-dimensional sensor data. To illustrate this procedure, before and after registration comparison for only five engines from the training data of sensor T24 (see Fig.\ref{fig:registrationbg})

\newpage

\subsection{From Discrete data to Multivariate Functional 
Data using Bspline Smoothing} \label{smoothing}

\begin{figure}[t]
\centering
\includegraphics[width=15cm]{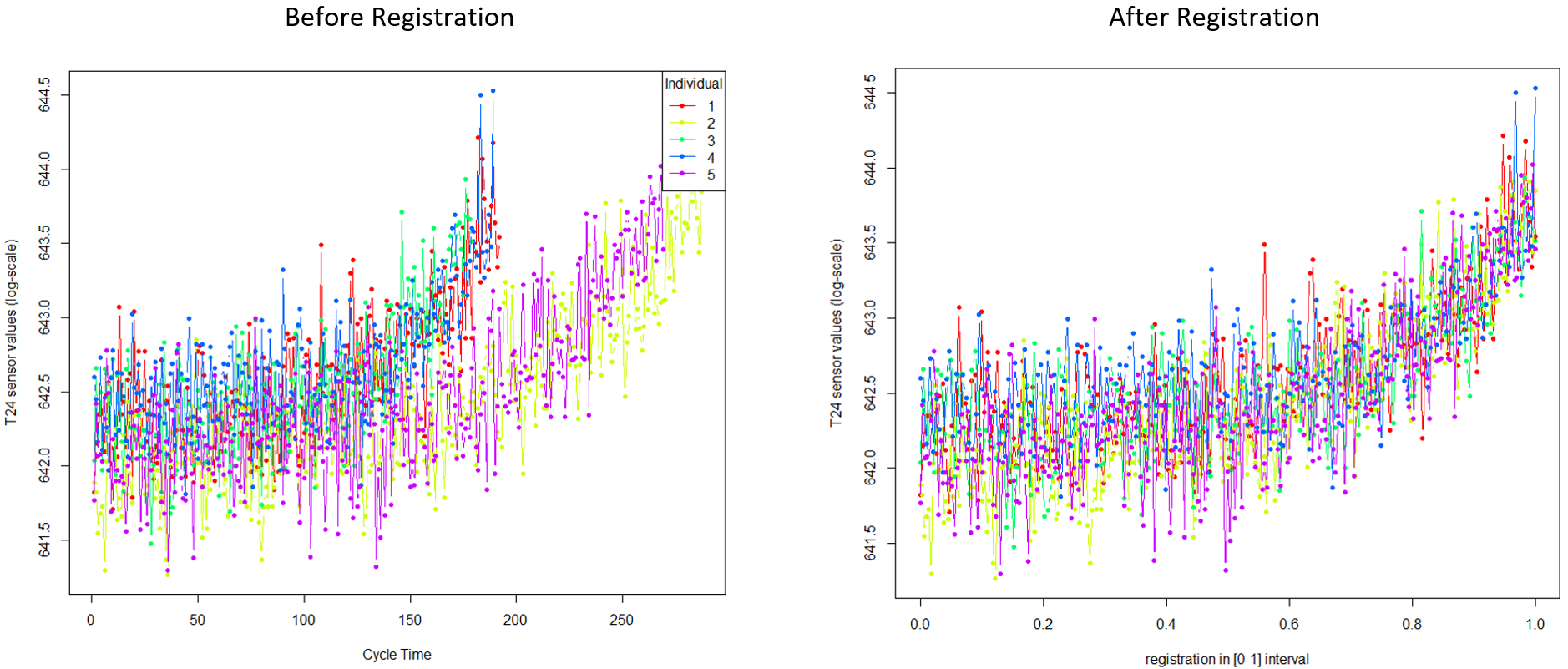}
    \caption{A comparison before and after registration for five random engines from Sensor T24}
    \label{fig:registrationbg}
\end{figure}

Similar to many financial and meteorological datasets, the engine sensor data in our study are recorded at discrete time points. Let $x_k$ denote an observed value of process $X(t)$ at the $k$th time point $T_k$ , where
$T$ is a compact set such that $t_k \in T$, for $k = 1, \mathellipsis, K$ . Then our data consists of $K$ pairs ($x_k, t_k$). These discrete data are smoothed to continuous functions in the form of Eq.\ref{eqXij}. as it is explained in Section \ref{prelim-smoothing}. This procedure is repeated for each sensor domain. Consequently, the final format of the equation of smooth sensor curves can be expressed as
\begin{equation}
X^*_{ij} = \displaystyle\sum_{b=1}^{B} c_{ijb}\phi_b(t), \quad i=1,\mathellipsis,n \quad j=1,\mathellipsis,J.
\end{equation}
where $c_{ijb}$ are the coefficients of the basis functions $\phi_b(t)$ for $i^{th}$ engine, $j^{th}$ sensor and $b^{th}$ basis function and $B$ is number of basis functions. Cubic Bsplines are used to smooth functions (\ref{cubicsplines}).  GCV is used to find an optimal number of splines and minimize the MSE (\ref{eqgcv}). Smooth functional data generated from discrete observations of a random engine for the T24 sensor can be found in Fig.\ref{fig:bspline2}.

\begin{figure}[ht]
    \centering
    \includegraphics[width=10cm]{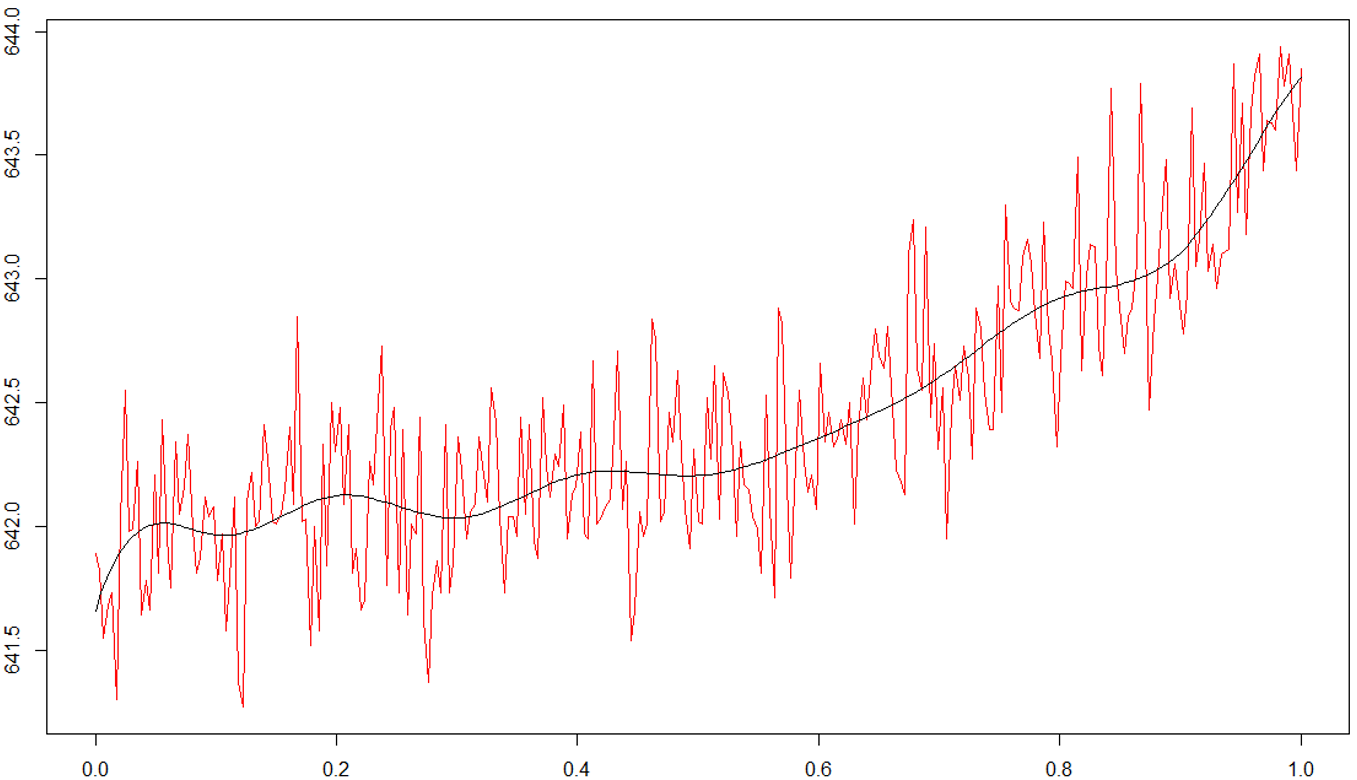}
    \caption{From the observed data to smooth functional data using cubic Bspline  after choosing optimal $\lambda$ with GCV.}
    \label{fig:bspline2}
\end{figure}

All 9 sensor curves after registration and smoothing are given in Fig.\ref{fig:smooth_reg}. Different colors in each graph represent the 100 individual engines from the training dataset.

\begin{figure}[ht]
    \centering
    \includegraphics[width=15cm]{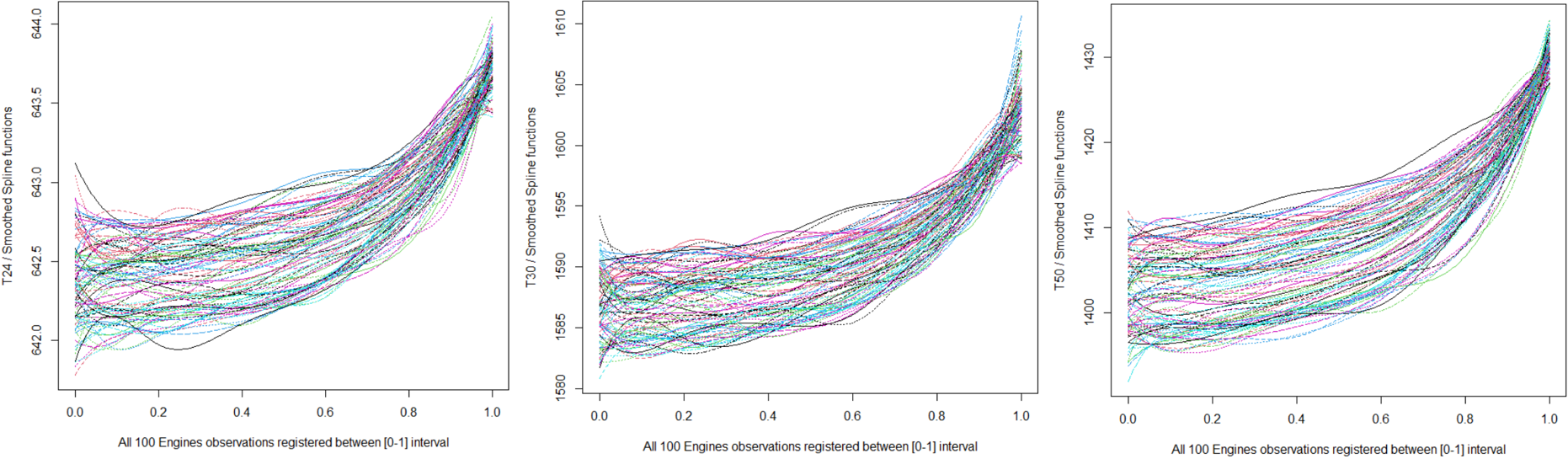}

    \centering
    \includegraphics[width=15cm]{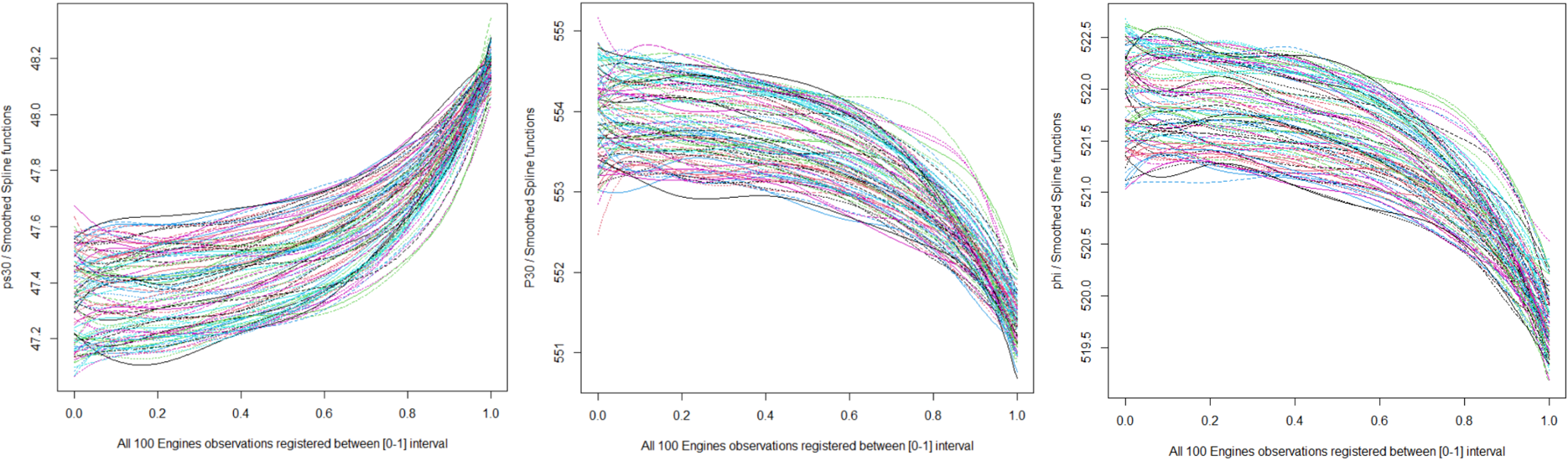}

    \centering
    \includegraphics[width=15cm]{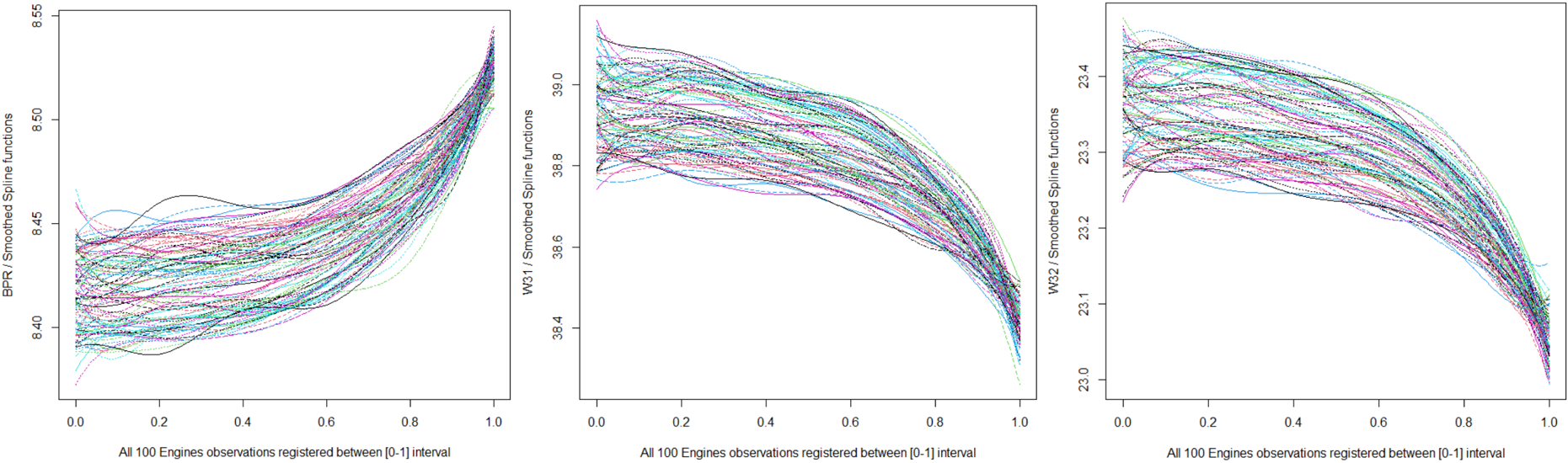}
    \caption{Smooth curves generated for the 9 informative sensors within the [0,1] interval}
    \label{fig:smooth_reg}
\end{figure}

\subsection{MFPCA and KLE Reconstruction} \label{MFPCAdataexplanation}

The failure behavior of an aircraft engine is a stochastic process affected by uncertainties arising from physical degradation dynamics, changes in usage, and other effects. The degradation pattern of an aircraft engine is complex and not fully understood. To overcome this complexity, MFPCA is applied to understand and explain the data in the multivariate sense for the training dataset. Estimation of $k$th principal component score of $i$th engine and $j$th sensor is expressed (from Eq.\ref{eqksi}) in multivariate sense as
\begin{equation}
\xi_{ijk}=\int^1_0 \big(X^*_{ij}(t)-\bar{X}^*_j(t)\big)         \widehat{f}^*_k(t)dt
\end{equation}
where $i=1,\mathellipsis,100$ (the number of engines), and $j=1,\mathellipsis,9$ (the number of informative sensors) and $\bar{X}^*_j(t)$ is the mean function for $j$th sensor expressed as
\begin{equation}
\bar{X}^*_j(t):= \frac{1}{n} \sum^n_{i=1} X^*_{ij}(t),
\end{equation}
and the weight function $\widehat{f}^*_k(t)$ is the eigenfunctions of covariance operator $\widehat{C}^*$. The solution of the second order eigenequation (see Eq.\ref{eqeigenvalues}) for each sensor is given as 
\begin{equation}
\widehat{C}_j^*(f^*_k)(t)=\int^1_0 \widehat{C}_j^*(t,u)\widehat{f}^*_k(u)du = \widehat{\lambda}^*_k \widehat{f}^*_k(t), \quad t \in I,
\end{equation}
where $\widehat{C}_j^*(t,u)$ is $j$th sensor's covariance  function is
\begin{equation}
\widehat{C}_j^*(t,u)=\frac{1}{n-1} \sum^n_{i=1}(X^*_{ij}(t) - \bar{X}^*_j(t))(X^*_{ij}(u) - \bar{X}^*_j(u)).
\end{equation}

The approximation of the curves truncating the KLE in terms of  the first $q$ principal components (see Eq.\ref{eqxqu})
\begin{equation}
X^{*^q}_{ij}(t)=\bar{X}^*_j(t) + \sum_{k=1}^q\widehat{\xi}^*_{ijk}\widehat{f}^*_k,
\end{equation}
and explained total variance is given by $\sum^q_{k=1}\widehat{\lambda}^*_k$  (\ref{eqtotalvariance}), can be found in Table \ref{tblpropvar}. The KLE reconstructions of the sensors ps30 and W31 with the first MFPC for the two random engines (82, 49) are given in Fig.\ref{fig:KLErecons}. Straight lines represent the real curve after smoothing, and the red dashed lines represent our reconstruction with KLE after MFPCA. KLE reconstructions verify how the first MFPC fits with the real sensor curves. The first two principal components explain 99.6 percent of the total variance, and only the first principal component already explains 96.4 percent of the total variance. The first three Multivariate Functional Principal Components (MFPCs) are given in Fig. \ref{fig:MPFCs}.
\begin{figure}[hb]
\centering
\includegraphics[width=15cm]{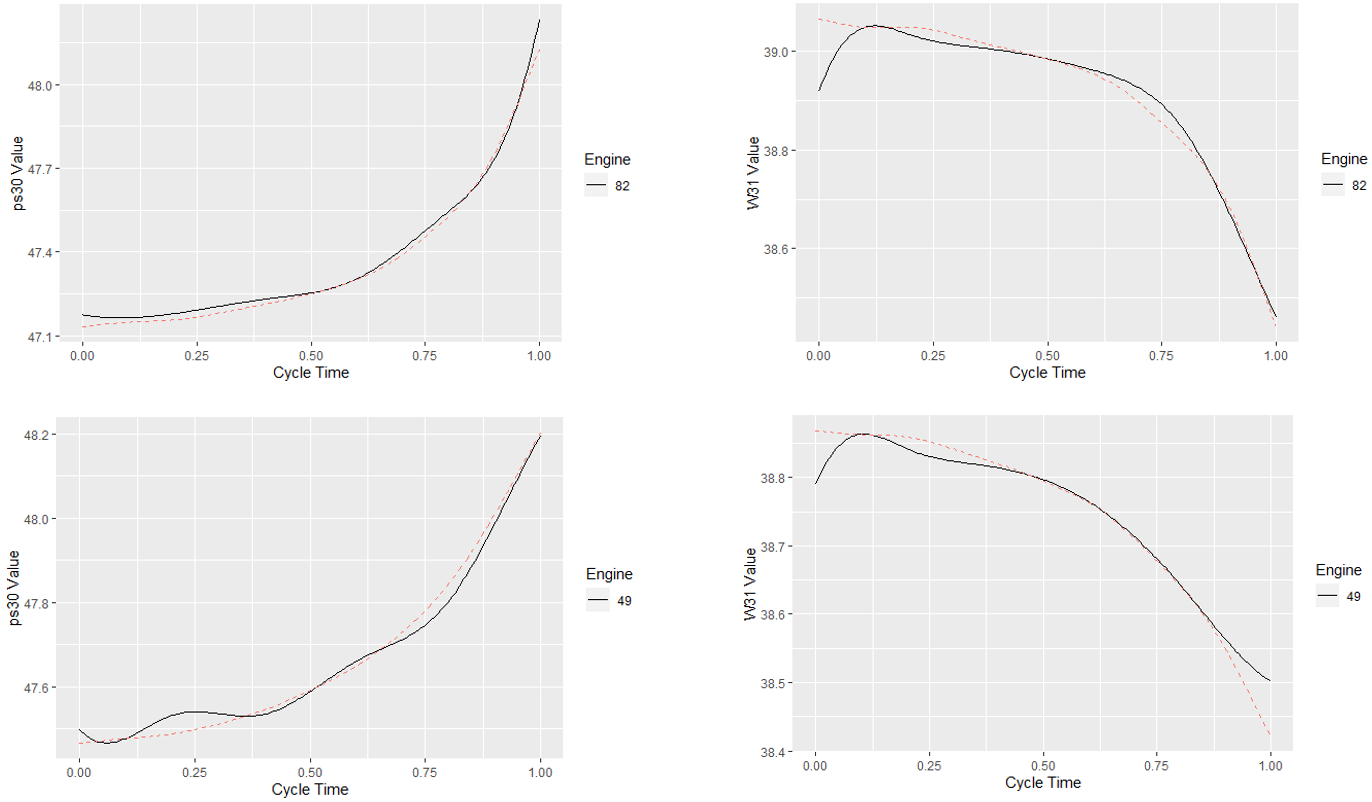}
    \caption{KLE reconstructions of two random engines for sensors ps30 and W31}
    \label{fig:KLErecons}
\end{figure}
\begin{figure}[t]
    \centering
    \includegraphics[width=15cm]{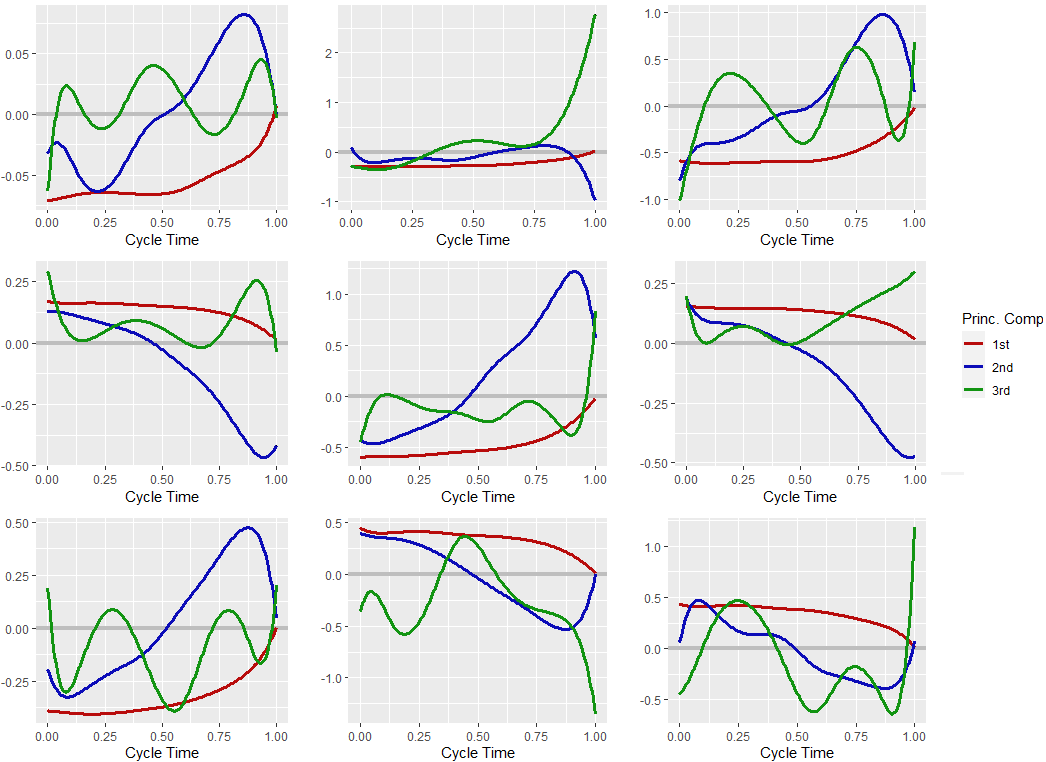}
    \caption{First 3 MFPCs for the 9 informative sensors}
    \label{fig:MPFCs}
\end{figure}

\begin{table}
\caption{Total Variance Explained by First 3 MFPCs}\label{tblpropvar}
\begin{tabular}{ll}
\hline\noalign{\smallskip}
Principal Components & Proportion of Variance Explained\\
\noalign{\smallskip}\hline\noalign{\smallskip}
1st PC & 96.4\\
2nd PC & 3.2\\
3nd PC & 0.4\\
\noalign{\smallskip}\hline
\end{tabular}
\end{table}

In our practical case study, the first multivariate functional principal component explains 96.4\% of the total variation, meaning that the first MFPC is highly informative. When exemining the MFPC scores of the first principal component for all 100 engines and nine informative sensors in the training dataset, we see that they can be modeled as a mixture of two normal distributions (see Fig.\ref{fig:scores_and_density}). In this case, it can be assumed that the low scores of the first MFPC fit the normal distribution with a mean of -2.874 and standard deviation of 2.649, while the high scores fit the normal distribution with a mean of 4.893 and standard deviation of 2.649.

\begin{figure}[t]
\includegraphics[width=7cm]{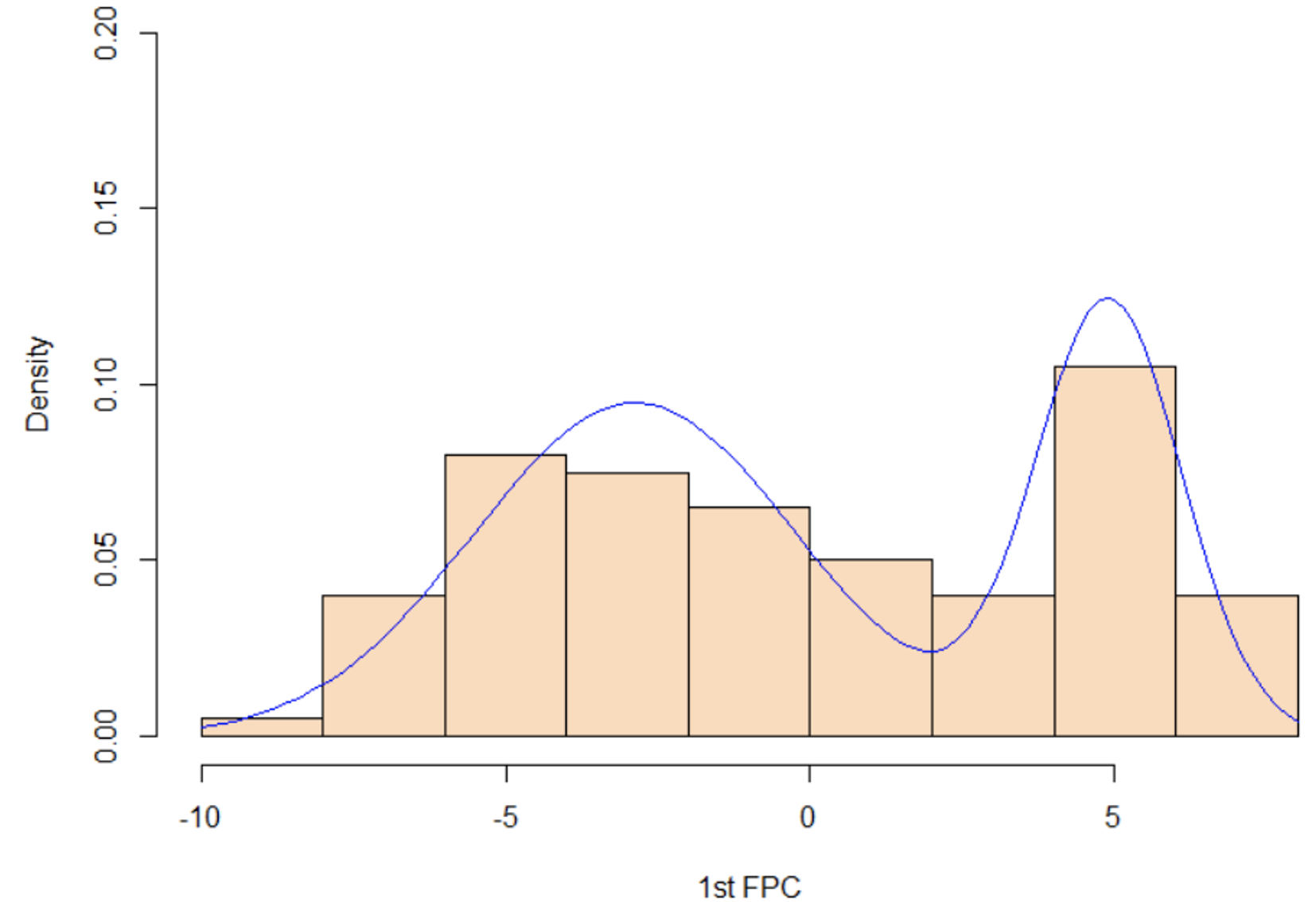}
    \caption{Histogram of the first MFPCs with a bimodal density function}
    \label{fig:scores_and_density}
\end{figure}

From Fig. \ref{fig:smooth_reg}, we can see that the sensors show two different possible behaviors: T24, T30, T50, phi, and BPR have an increasing trend, and P30, ps30, W31, and W32 have a decreasing trend. To represent these two different trends, Fig.\ref{fig:redblackregistered} gives all curves for only sensors, T24 (left graph), and W32 (right graph); red-colored engines indicate the low-scoring group and black-colored ones indicate the high-scoring group. The yellow line is the mean function of the black group, and the green line is the mean function of the red group.
The yellow line represents the mean function of the black group, while the green line represents the mean function of the red group. The mean functions are quite different for each group, as shown in Figure \ref{fig:redblackregistered}. It is evident that the initial point plays a decisive role in separating the two groups. Essentially, the interpretation is that if the initial MFPC score is low or high, the starting point will have correspondingly lower or higher values. As discussed \cite{saxena2008damage}, each engine operates with unknown varying degrees of initial wear and manufacturing variation. Therefore, another interpretation of the starting points of the sensor values and the first principal component scores could be related to that wear and manufacturing variation. On the other hand, the same training samples are also investigated on the original scale (without registration) in Fig.\ref{fig:redblackorj}. Moreover, it can be seen that the starting point of each group is different, and the total lifespan is also different. The black (high score) group lives longer than the red one. Now, similar failure behavior is expected to be obtained from the test dataset, and we will use all the information we learned from the training dataset. This information, which allows us to divide the test dataset into two groups, will be useful for RUL estimation of the test dataset.
\begin{figure*}[b]
\centering
\includegraphics[width=14.2cm]{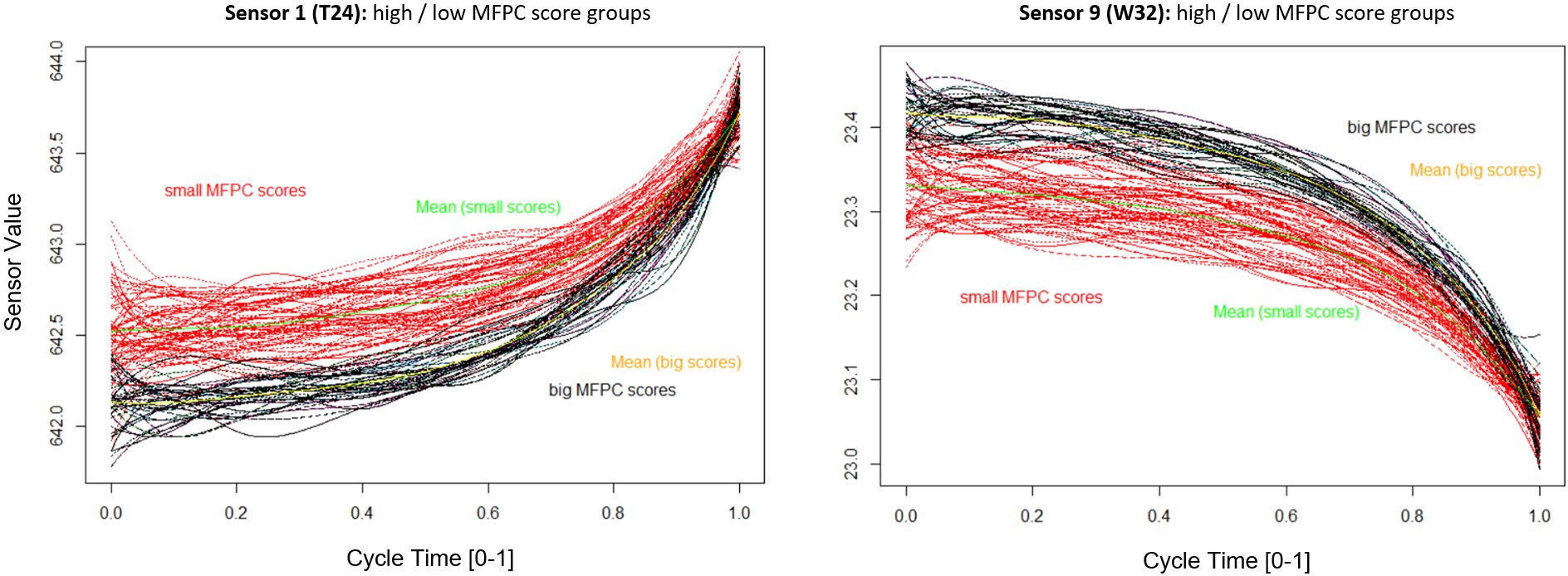}
    \caption{Classification of small and big scores for 100 training engines for one sensor with an increasing trend over time (T24) and one sensor with a decreasing trend over time (W32) in the registered interval [0,1]}
    \label{fig:redblackregistered}

\includegraphics[width=14.2cm]{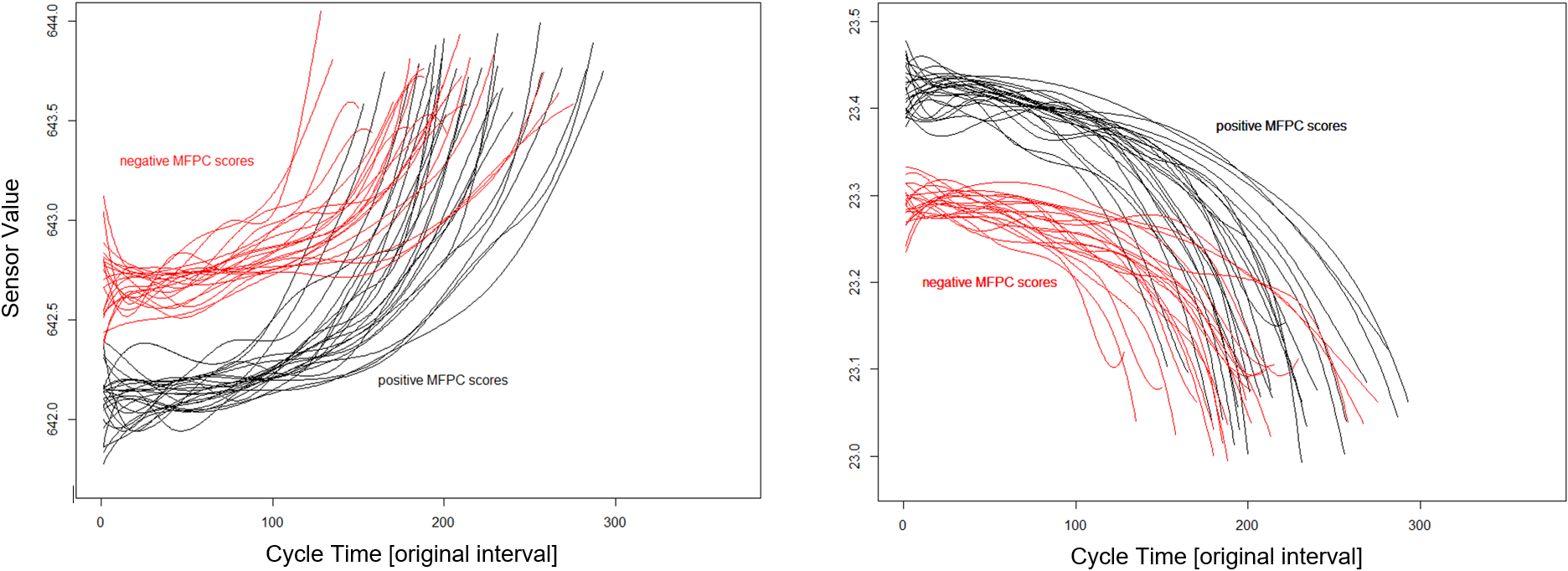}
    \caption {Classification of small and big scores for 100 training engines for one sensor with an increasing trend over time (T24) and one sensor with a decreasing trend over time (W32) in original scale}
    \label{fig:redblackorj}
\end{figure*}





\subsection{Estimation of the Sensors' Behavior during the Degradation Process and the RUL Prediction} \label{distancecalculation}

\begin{figure}[t]
\centering
\includegraphics[width=15cm]{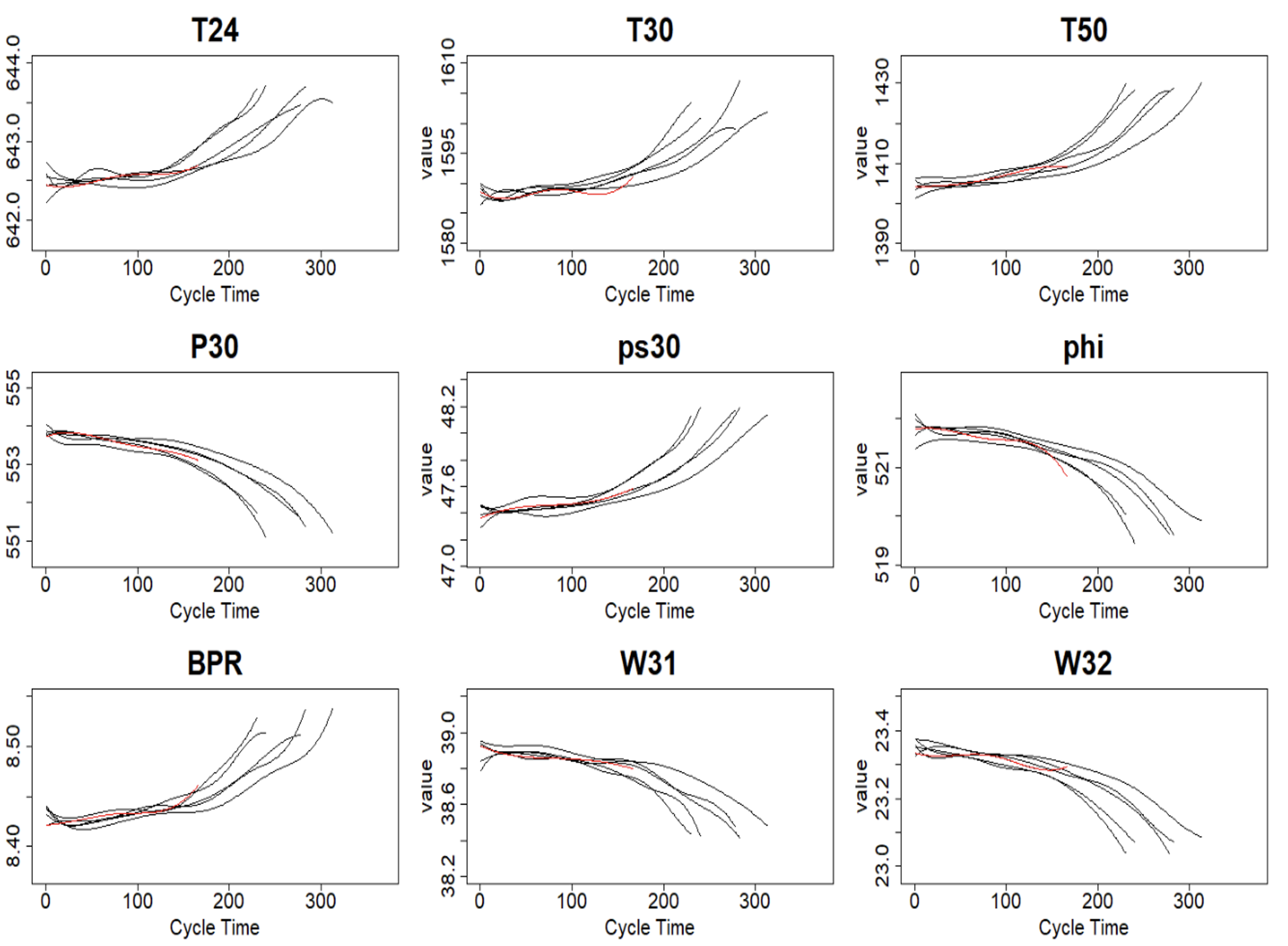}
    \caption{Test engine number 20 and its closest training engines after multivariate distance calculation (the graph is provided for all 9 informative sensors)}
    \label{fig:9dimclosesttest20}
\end{figure}
The next step of the proposed method (see Fig. \ref{fig:steps1}.) involves applying a similarity-based method to determine which test engine is similar to which training engine after learning the degradation behavior of the training dataset. Once the MFPC score classification is done in Section \ref{MFPCAdataexplanation}, we can calculate the distance between a test engine curve and all training engine curves in the same group. Moreover, we can select the closest training engines that can be defined as the most similar. However, distance is calculated only with training engines that have outlived the relevant test engine. In other words, training engines that fail earlier than the relevant test engine must be eliminated.

Using the equation (\ref{eq:distance1}.) given in Section \ref{Similaritybased}, the sensor value is normalized, and the multivariate distance between the relevant test engine and all training engines is calculated (\ref{eq:distance2}). The distances between each test engine and all training engines in the same MFPC Score group are calculated after the classification of the test engines. Therefore, grouping the test dataset is also done here. To do this, the initial starting point for each sensor is considered in the training dataset, and this information is used to classify groups in the test dataset. Youden Index \cite{youden1950index} is used to provide the optimal cutoff point that will allow us to classify a new test engine into one of the two populations. The Youden index is a single statistic that captures the performance of a dichotomous test. It has been successfully applied in many medical studies (e.g., \cite{aoki1997evaluation, castle2003comparison, fluss2005estimation}) to provide a convenient one-dimensional summary of test accuracy and determine its associated cutoff point. For example, if a test engine is classified in the black group, the distance between this engine and all black group engines in the training dataset is calculated.  Therefore, the failure time of this engine is expected to be similar to training engines. Figure \ref{fig:9dimclosesttest20}. gives an example of test engine 20 and the closest training engines after multivariate distance calculation. The red engine in the figure shows the test engine that has not failed yet, and the black engines are the five most similar engines that failed in the training dataset. Using these similar training engines, RUL prediction and estimated multi-sensor curves during the RUL will be obtained.

The mean of the failure times of the most similar training engines and the median of the failure times of the most similar training engines are separately used to predict the RUL. For example, when we checked the test engine 20 given in Fig.\ref{fig:9dimclosesttest20}, we learned that the test engine (red color) would behave like these training engines (black color) and was expected to fail at a similar point. Not only can the RUL of each engine be estimated, but the multi-sensor curve to the predicted point of failure can also be estimated using the mean of sensor values of similar training engine curves at relevant time points. These curve predictions are performed for all 100 test engines and all nine informative sensors. An example of the entire engine curves estimate for T24 is given in Fig.\ref{fig:alltestpred}. Here, the black and red MFPC groups are also shown separately in the figure.
\begin{figure}[t]
\centering
\includegraphics[width=8cm]{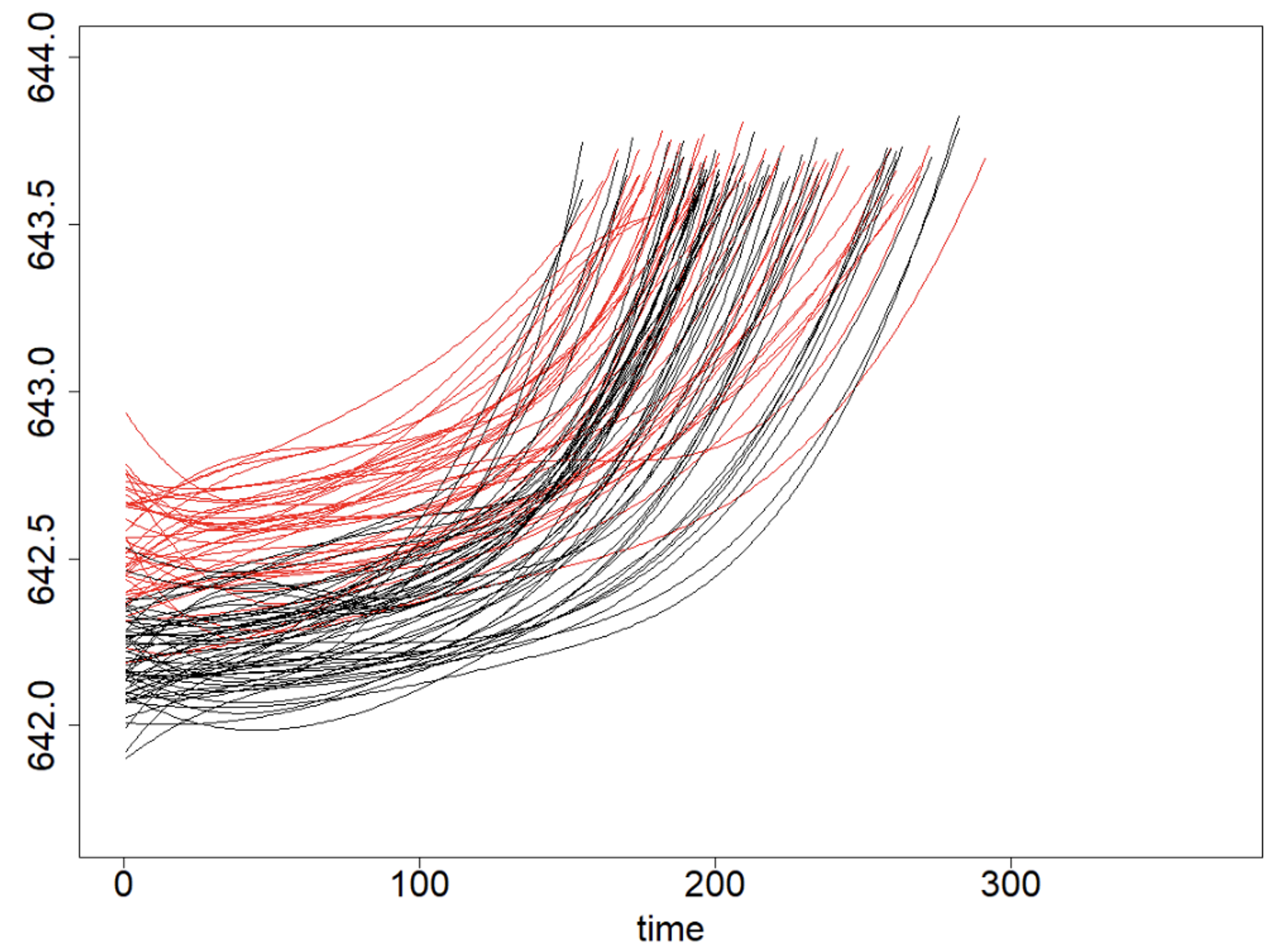}
    \caption{All test engines curve predictions for sensor T24}
    \label{fig:alltestpred}
\end{figure}

To illustrate the importance of sensor curve estimation and how the predicted curves fit the actual curve, two examples of predicted curves for test engines 10 and 70 are given in Figure \ref{fig:curve10}. While the blue curves in the relevant figures show the total observations made so far for the relevant engines, the black curve gives our curve estimate during the predicted RUL.
\begin{figure}[ht]
\centering
\includegraphics[width=15cm]{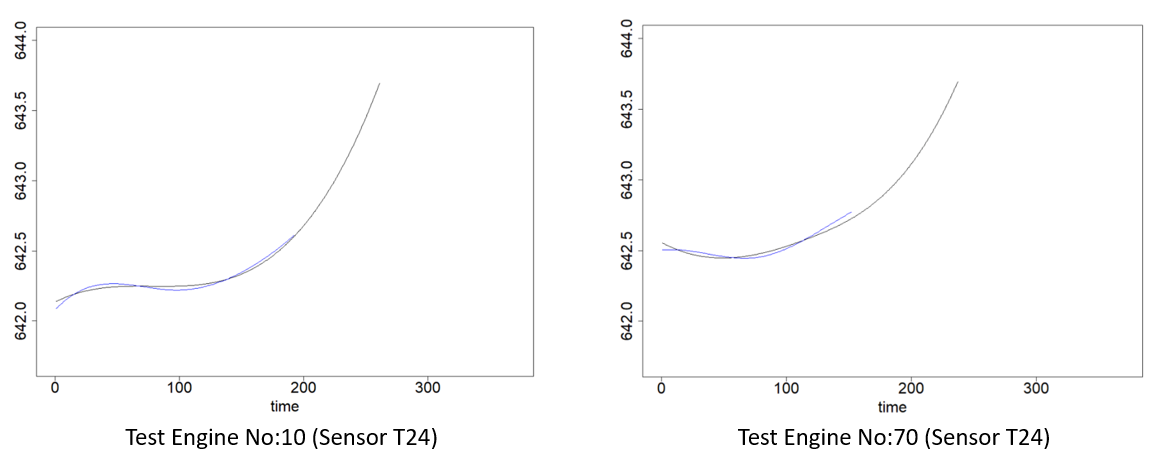}
    \caption{Test engine number 10 and test engine number 70 curve estimations during the predicted RUL for the sensor T24}
    \label{fig:curve10}
\end{figure}

\subsection{Results and the Interpretations} \label{rulpredsection}

The common indicator ``Root Mean Square Error (RMSE)'', was used to evaluate the performance of the prognostic method for the C-MAPSS dataset. However, one of the challenges of such problems was identified as the interpretability of the RUL found \cite{vollert2021challenges}. Therefore, the main motivation of the paper is not only the RUL prediction performance but also the interpretation of sensor curves during RUL.

\begin{figure}[t]
    \centering
    \includegraphics[width=15cm]{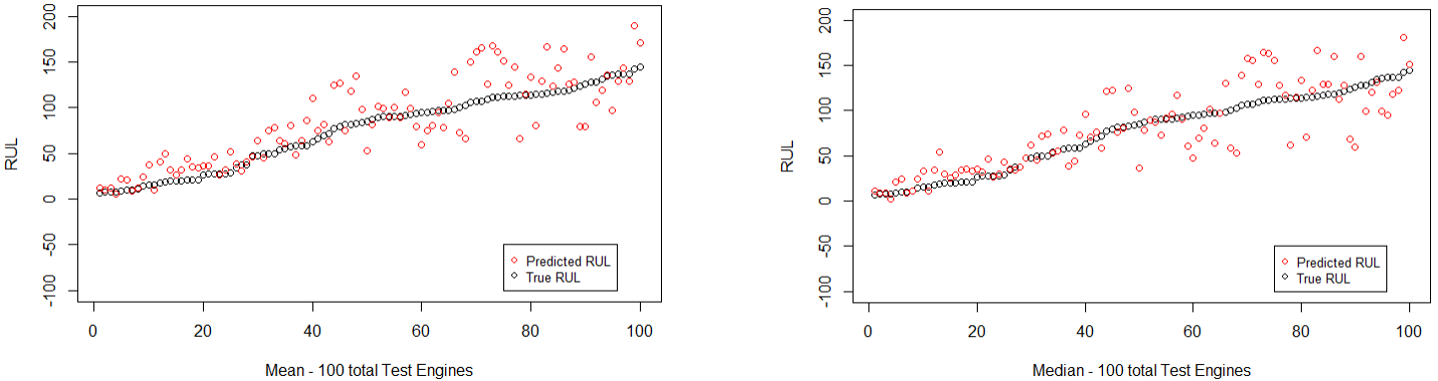}
    \caption{Predicted and true RUL using mean and median of the closest training engines, ordered by increasing RUL}
    \label{fig:meanpred}
\end{figure}

The number of the most similar engines to consider (see Section \ref{distancecalculation}.) is investigated through sensitivity analysis to analyze how the choice of the number of similar training engines will affect the methodology in terms of RMSE. Similar results were found, but after sensitivity analysis, taking the closest six or eight engines yielded the best results. RUL prediction using mean and median, and the true RULs with an increasing order are given in Fig.\ref{fig:meanpred}.  Although there are better RMSE results obtained using different ML approaches in the literature for the given dataset, the proposed novel multivariate FDA approach also provides competitive results in terms of RMSE. Especially compared with other statistical and univariate FDA approaches  \cite{zhang2020aircraft} in the literature (see Table \ref{tbl3}.), the RMSE (both mean and median) of the proposed prognostic method is lower, indicating that the performance is better than the FPCA model and Exp model. In addition, a narrower prediction error range and a higher number of correct predictions are obtained by using the proposed method. It is shown from both mean and median predictions that the prediction error is lower than expected when the RUL is lower. As a result, since it considers raw multi-sensor data instead of extracted univariate health index (HI), the MFPCA predicted the RUL better than the univariate FPCA.

\begin{table}
\caption{Performance comparison}\label{tbl3}
\begin{tabular}{llll}
\hline\noalign{\smallskip}
\textbf{Method}& \textbf{RMSE} & \textbf{Range of Prediction Errors}&  \textbf{Correct Pred.}\\   
\noalign{\smallskip}\hline\noalign{\smallskip}
Exp  & 45.40 & [-135,63]  & 1\\
FPCA (Mean)  & 28.06  & [-82,65]  & 3 \\
FPCA (Median)  & 28.70  & [-83,68]  & 5 \\
M-FPCA / Proposed Method (Mean)  & 25.41  & [-57,58]  & 7  \\
M-FPCA / Proposed Method (Median) & 25.74  & [-66,53]  & 1  \\
\noalign{\smallskip}\hline
\end{tabular}
\end{table}

One of the most important benefits of the proposed method can be defined as the interpretability of raw sensor curves on RUL. Curve prediction was very important because it would give us input into the degradation process. Sensor curve estimation is done after the end of the observations for each test engine. However, the RMSE performance of the curves was also investigated and compared when we predicted the curves at an earlier time such as 35, 50, 80, and 90 percent of the failure time for the test engines. This study was conducted for test engines 20, 31, 34, 35, 42, 68, 76, 81, and 82 to compare the results with a univariate FPCA study \cite{zhang2020aircraft} done, which compares the univariate FPCA with Exp model. The RMSE of the predicted sensor curves was found to be lower than that of univariate FPCA. An example for test engine number 82 is given in Fig.\ref{fig:univsmulti}. In these graphs, blue curves represent the given percentage of observations, red curves represent the curve prediction after the applied method, and black dots represent the actual observed sensor values. Note that this engine is from the test dataset and has not failed yet. Fig.\ref{fig:comparisontable} represents the RMSE ratio of the curves for MFPCA and Uni-FPCA, and the RMSE average of 9 sensors is taken into account to obtain an overall comparison table. The fact that the MFPCA/uni-FPCA RMSE ratio is less than 1 proves that MFPCA predicts the sensor curves better.

As discussed in Section \ref{Similaritybased}, one of the key advantages of the FDA is its use of derivatives of functions. When we check the 2\textsuperscript{nd} derivatives of the high and low MFPC score groups found in Section \ref{MFPCAdataexplanation}, the 2\textsuperscript{nd} derivatives appear to have a dramatic increasing or decreasing behavior across different sensors after 80 percent of the total life. The mean function of 2\textsuperscript{nd} derivatives for both the black and red groups is shown in Fig.\ref{fig:deriv2nd}. Such information is not available in most ML approaches that focus solely on RUL performance in PHM. This information can be used as an alarm point for activities such as maintenance decisions and replacements. As can be seen from Table \ref{tbl5}., when we place an alarm into the 80\% percent of the predicted RUL, 95\% of the engines have a maintenance alarm before the actual failure time. The remaining 5\%  percent shows the maintenance alarm further ahead of the real failure point. However, when these engines are examined in detail, it is seen that they have very few observations yet. In other words, the engines are still at the beginning of their lives, with very little observation. Since the alarm point will change dynamically as the number of observations increases, the interpretation will also change dynamically over time, and these engines are expected to alarm before the actual fault point in the future. Also, one of the essential considerations in the dynamical interpretation can be described as "too early maintenance." Besides RUL performance and RUL interpretability, one of the most important challenges of PHM approaches is given in the literature as implementability. Thanks to the dynamic sensor curves estimated in the proposed FDA approach, the alarm point will also be shifted dynamically over time. In this way, as the number of observations increases, the alarm point can be estimated as more stable, which is very important for the implementation of an effective condition monitoring system. If we check Table \ref{tbl5}, 58 percent of the maintenance alarms are triggered in the remaining 20 percent of total life, and 94 percent of the alarms are triggered in the last 40 percent of total life. This means that the proposed method can not only competitively predict RUL, but also estimate and interpret the sensor curves and implement dynamic alarms before the failure occurs. At the same time, it does not trigger the alarm too early, resulting in a lack of availability problems.
\begin{figure}[t]
  \centering
    \includegraphics[width=15.4cm]{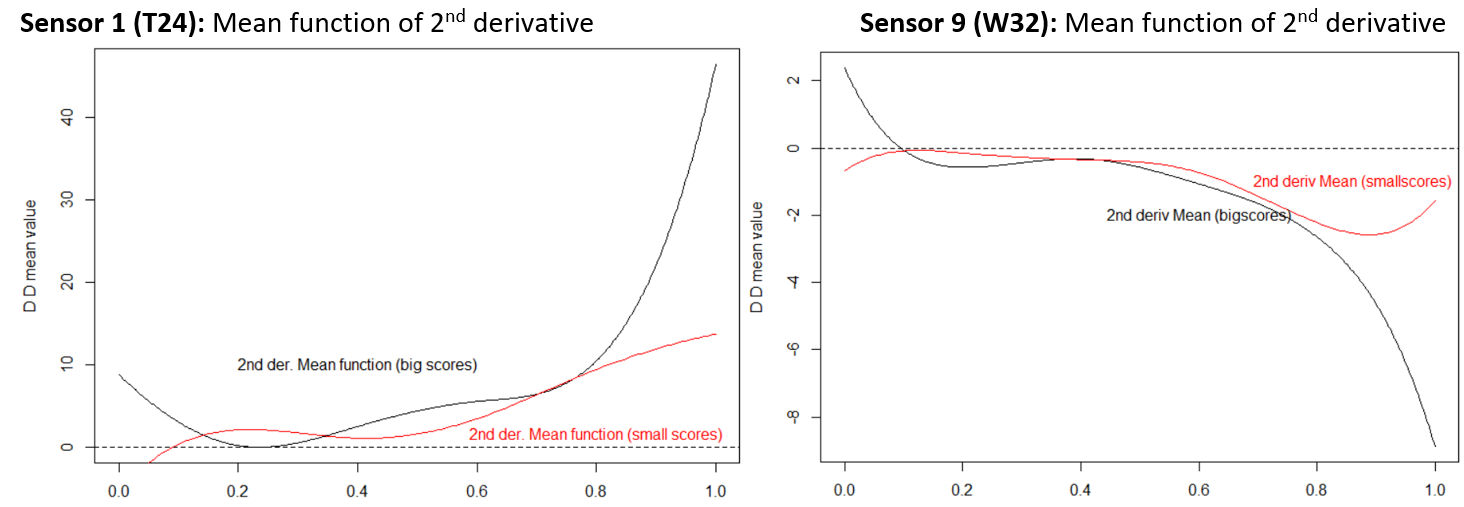}
    \caption{The mean functions of the 2\textsuperscript{nd} derivatives of the high-score and the low-score classified groups (for sensors T24 and W32, as shown in the figure) }
    \label{fig:deriv2nd}
\end{figure}

\begin{figure*}[p]
\centering
\includegraphics[width=11cm]{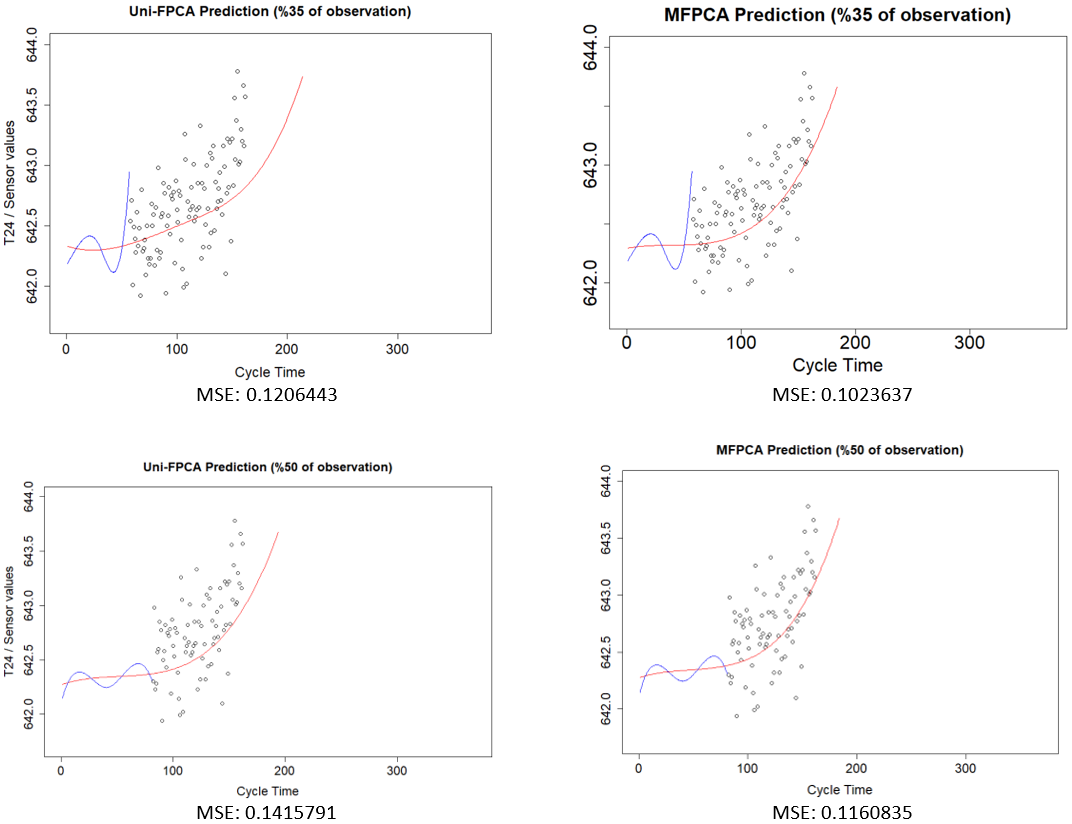}

\includegraphics[width=11cm]{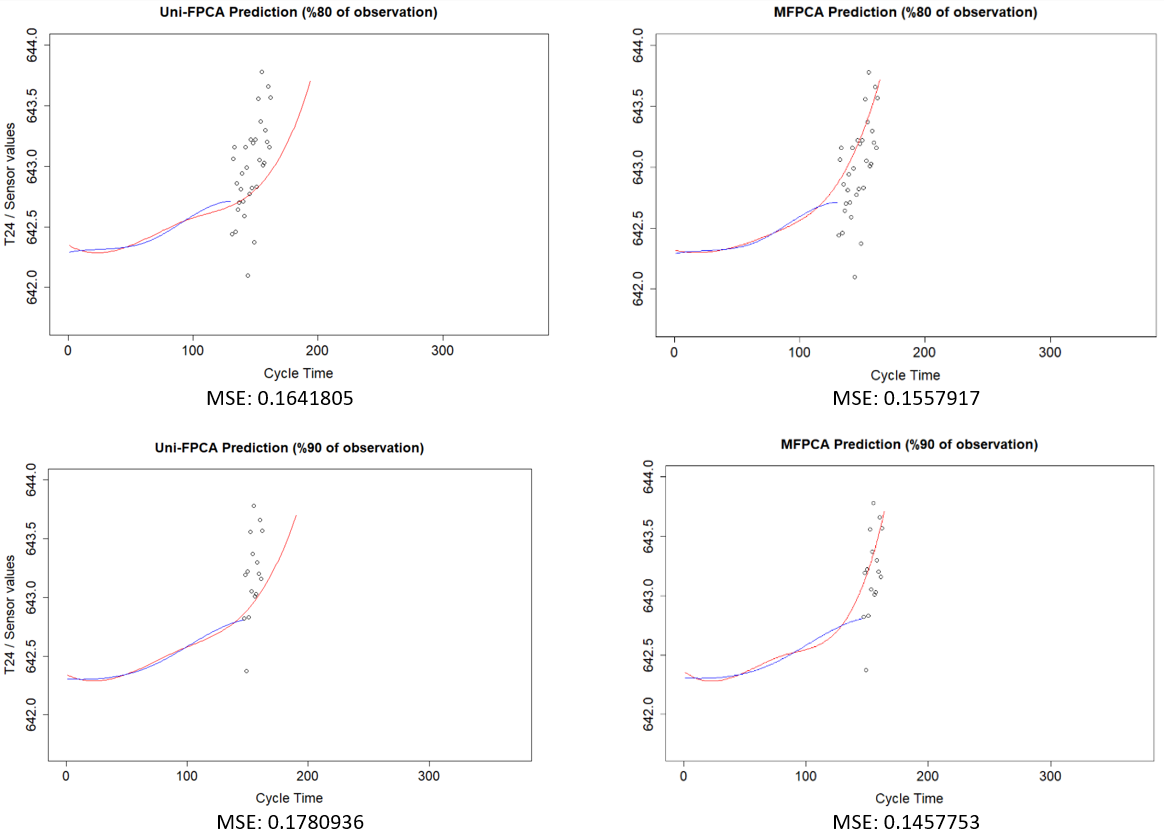}
    \caption{Comparison of curve prediction after Univariate FPCA and Multivariate FPCA for test engine 82 (sensor T24).}
    \label{fig:univsmulti}

\includegraphics[width=9.6cm]{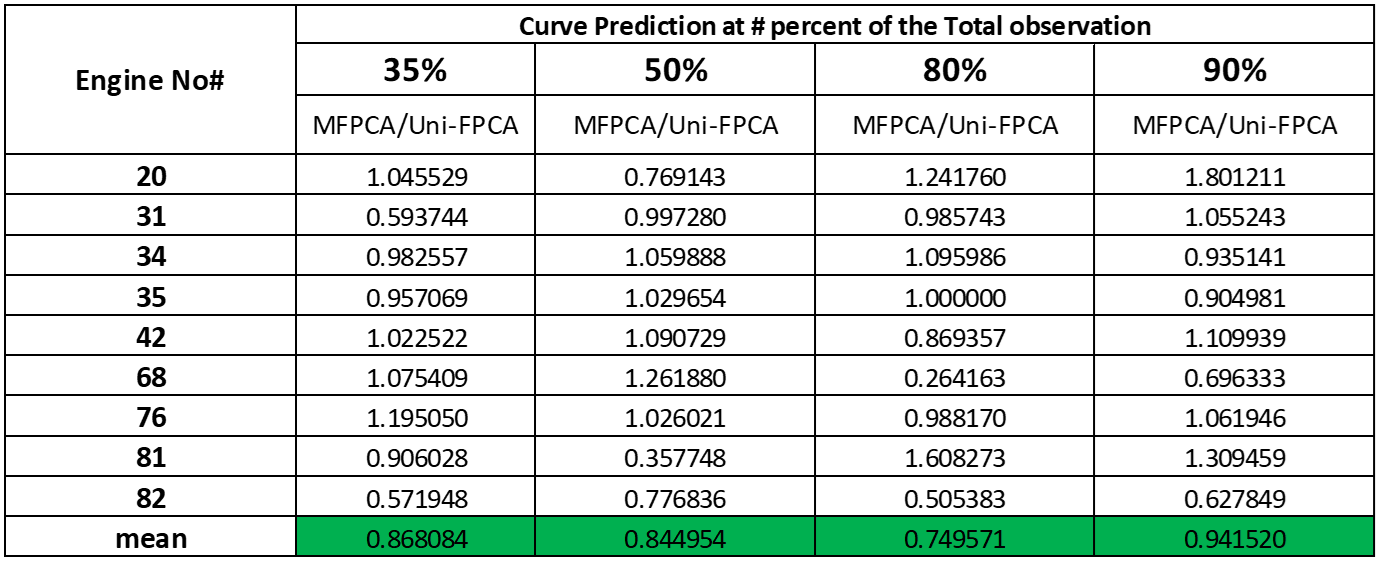}
    \caption{Comparison of curve prediction RMSE between Univariate FPCA and Multivariate FPCA for several test engines, with the MFPCA/Uni-FPCA RMSE ratio provided for the average of all sensors in a multivariate context.}
    \label{fig:comparisontable}
\end{figure*}

\begin{table}
\caption{Alarm Points Performance}
\label{tbl5}
\begin{tabular}{ll}
\hline\noalign{\smallskip}
  \textbf{} & \textbf{Unit}\\    
\noalign{\smallskip}\hline\noalign{\smallskip}
No of Total Test Engines &  100\\
"Alarm Point" is later than "True Failure Point" \quad   & 5\\
"Alarm Point" is earlier than "True Failure Point" \quad  & 95\\
"Alarm Point" is in last 40\% of Total Life & 94\\
"Alarm Point" is in last 30\% of Total Life & 87\\
"Alarm Point" is in last 20\% of Total Life & 58\\
"Alarm Point" is in last 10\% of Total Life & 22\\
"Alarm Point" is in last 5\% of Total Life & 8 \\
\noalign{\smallskip}\hline
\end{tabular}
\end{table}

\section{Conclusions} \label{Conclusions}

With the growing interest in the fields of IoT, ML, and Industry 4.0 applications, PHM in the aviation industry became important in reliability, safety, and maintainability engineering applications. In this paper, a new methodology is presented for RUL prediction and interpreting RUL post-prediction, addressing one of the biggest challenges in PHM. The benefits of a novel multivariate FDA approach are demonstrated considering some key challenges in PHM studies, with results compared to existingliterature.

Engine degradation data, divided into training and testing data, is first pre-processed to make it suitable for an MFPCA application. For this purpose, well-known FDA tools such as registration and smoothing with Bsplines are used in the pre-processing steps. MFPCA is then applied to learn the degradation behavior of the systems. After learning the degradation behavior of the sensors, the distance between the test and training multi-sensor curves is taken into account to find the most similar training engines for each test engine. Instead of feature extraction and univariate health index (HI) generation, which can lead to information loss, raw multisensor data is directly taken into consideration. In addition, one of the most important advantages of the FDA, the interpretation of derivative functions, is shown. The results obtained from the study show that the proposed method estimates RUL in a competitive place compared to other studies in the literature, estimates RUL better than uni-FPCA studies in the literature, and also helps in making dynamic alarm decisions thanks to the RUL interpretation. Furthermore, the study provides good practices to prove the eligibility of using multivariate FDA techniques in related areas. A novel, step-by-step approach is proposed as a guide for the field of PHM.

Some possible extensions of this work could involve using multivariate FDA not only for RUL prediction and interpretation but also for other PHM applications such as condition monitoring, fault classification, real-time monitoring, and predictive maintenance.

\section*{Conflict of interest}
The authors declare that they have no conflict of interest.


%
%

\bibliographystyle{spmpsci}
\bibliography{mybib}


\end{document}